\documentclass[showpacs,preprintnumbers,amsmath,amssymb]{revtex4}
\usepackage{amsmath,amssymb,graphics,epsfig,subfigure}
\usepackage{color}

\begin{document}
\renewcommand{\baselinestretch}{1.3}

\title{Triple points and novel phase transitions of dyonic AdS black holes with quasitopological electromagnetism}
	
\author{Ming-Da Li$^1$$^2$, Hui-Min Wang$^1$$^2$, Shao-Wen Wei$^1$$^2$ \footnote{Corresponding author. weishw@lzu.edu.cn}}
	
\affiliation{$^{1}$Lanzhou Center for Theoretical Physics, Key Laboratory of Theoretical Physics of Gansu Province, School of Physical Science and Technology, Lanzhou University, Lanzhou 730000, China\\
$^{2}$Institute of Theoretical Physics and Research Center of Gravitation,
Lanzhou University, Lanzhou 730000, People's Republic of China}

\begin{abstract}
Quasitopological electromagnetism has an important influence on the strong gravity of dyonic black hole. For example three photon spheres with one being stable are found and we wish to test them with the observed black hole shadow. We are primarily concerned with the thermodynamics and phase transition for a dyonic anti-de Sitter (AdS) black hole when the quasitopological electromagnetism is included.Unlike in a black hole solution without quasitopological electromagnetism, we observe a triple point phase structure by varying the coupling parameters. Of particular interest, for certain parameter values, two separate coexistence curves are present, which is an additional novel phase structure that is absent in a black hole solution without quasitopological electromagnetism. The critical exponents share the same values as mean field theory. These results uncover the intriguing properties of dyonic AdS black holes with quasitopological electromagnetism from a thermodynamic point of view.
\end{abstract}



\maketitle

\section{Introduction}

Thermodynamics and phase transition have composed one of the most active fields in the study of black holes. This suggests that black holes not only have the standard thermodynamic variables, such as temperature and entropy~\cite{BardeenHawkingCarter1973,Bekenstein1973, Hawking1975}, but also demonstrate a rich phase transition and phase diagrams. The pioneering work of Hawking and Page stated that there is a first-order black hole phase transition between large stable black holes and thermal radiation in anti-de Sitter space~\cite{HawkingPage1983}. This work opened the study of thermodynamic phase transition in black holes. Inspired by the anti-de Sitter/conformal field theory (AdS/CFT) literature~\cite{Maldacena1998,Gubser1998,Witten1998-1}, the Hawking-Page phase transition is interpreted as the gravitational dual of the confinement/deconfinement phase transition of gauge fields~\cite{Witten1998-2}. The study was also generalized to the charged and rotating AdS black holes~\cite{Chamblin,Chamblin2,Caldarelli1999}.

Recently, the thermodynamics of AdS black holes has been generalized to the extended phase space, where the cosmological constant was treated as thermodynamic pressure~\cite{Hen1984, Tei1985, Kastor2009, Dolan2011, Cvetic2011}. It was soon realized that a small-large black hole phase transition of the first order occurs in four-dimensional Reissner-Nordstr\"{o}m (RN) AdS black holes, which is analogous to a liquid/gas phase transition of the van der Waals (vdW) fluid. They also share the same critical phenomena~\cite{Kubiznak2012}. Subsequently, different typical phase transitions and phase structures including the reentrant phase transition, the triple point, and the superfluid phase were discovered in the extended phase space~\cite{Gunasekaran2012, Hendi2012, Chen2013, Altamirano2013, Cai2013, Altamirano2014, Mo2014, Altamirano2014-2, Frassino2014, Zou2014, Zou2014-2, Wei2014, Mo2014-2, Zhang2014, Pope2014, Kostouki2014, Hennigar2015, Wei2016, Hendi2016, Hendi2017, Hennigar2017, ZZY2017, Momeni2017, Chakraborty2015, Wei20152016, ZhangZouZhang2020,YZWL21} (for a recent review, see~\cite{MannTeo2017} and references therein). A study of the black hole phase transition was also applied to a test of the black hole microstructure. Novel dominated attractive and repulsive interactions have been discovered~\cite{Ruppeiner1979, Ruppeiner1995, Wei20152016, Wei2019-1, Wei2019-2,Yang:2018ixs}.

The study of thermodynamics indicates that the small-large black hole phase transition is universal for the charged AdS black holes in general relativity, modified gravity, or even high dimensions. Some phase transitions beyond the vdW-like one were also discovered. For example, the reentrant phase transition can be found in four-dimensional Born-Infeld AdS black hole coupling with a nonlinear electrodynamics~\cite{Gunasekaran2012}, and triple point in six-dimensional charged Gauss-Bonnet black holes~\cite{Wei2014, Wei2021}.

Quite recently, a black hole solution using a quasitopological electromagnetism has attracted great interest~\cite{Liu2020}. This theory is a new higher-order extension constructed with the bilinear norm of Maxwell's theory. When they are combined with global polarization, it was found that these additional terms do not contribute to the Maxwell equation and energy-momentum tensor. This is the reason why the theory is called quasitopological electromagnetism. Although quasitopological term makes no contribution to the purely electric or magnetic RN black holes, it indeed affects dyonic black hole solutions. For  black hole solutions with quasitopological electromagnetism~\cite{Liu2020}, it was found that the dominant, null, and the weak energy conditions are satisfied, while the strong energy condition can be violated in the matter sector. This modified black hole solution exhibits many novel features. In certain parameter regions, this black hole solution can possess four horizons and three photon spheres with one being stable. It also provides us with a nontrivial test of the topological charge of the black hole photon sphere~\cite{Wei2020}. In odd dimensions, the black hole solution was also discovered in Ref.~\cite{Cisterna2020}. There may be five black hole phases, where two of them are unstable while others are stable. Chaotic behaviors involving particle motion were explored in Ref.~\cite{Lei2021}. The results indicated that the chaos bound will be violated for this quasitopological electromagnetism. Scalarized black holes in the Einstein-Maxwell-scalar theory with a quasitopological term were constructed in Ref.~\cite{Myung2021}. Interestingly, it was found that the fundamental black hole is stable, but these excited ones are not. Black strings were also obtained in Ref.~\cite{Cisterna2021} in Einstein and Lovelock gravities. These studies shed new light on understanding the black hole solution with quasitopological electromagnetism.

Although dyonic black hole solutions with quasitopological electromagnetism exhibit some interesting results, the thermodynamic phase transition and phase diagram still remain to be studied. Since in some parameter regions there are four horizons and more than one photon sphere, one expects phase transitions beyond the vdW-like type may exist, which reveals the particularly interesting properties of black holes with quasitopological electromagnetism. Motivated by this, in this paper focus mainly om its thermodynamics, especially in the extended phase space.

The outline of this paper is as follows. In Sec.~\ref{s2}, we review the dyonic AdS black hole solution and give the corresponding thermodynamic quantities in four-dimensional Einstein gravity minimally coupling to standard Maxwell electromagnetism and quasitopological electromagnetism. In Sec.~\ref{s3}, we investigate the phase transitions and critical behaviors of dyonic AdS black holes by studying the characteristic behaviors of the temperature and the Gibbs free energy. Moreover, we depict the coexistence curve in $P$-$T$ diagram. In Sec.~\ref{s4}, we calculate the critical exponents near the critical points. Section.~\ref{s5} is devoted to conclusions and discussions.

\section{Thermodynamics of dyonic black holes}
\label{s2}

The Lagrangian of four-dimensional Einstein gravity minimally coupled to the Maxwell electromagnetism and quasitopological electromagnetism is
\begin{eqnarray} \mathcal{L}=\sqrt{-g}(R-2\Lambda)+\alpha_1\mathcal{L}_{\text{M}}+\alpha_2\mathcal{L}_{\text{Q}},
\end{eqnarray}
where negative $\Lambda$ is the cosmological constant in an AdS space. The parameter $\alpha_1$ is a dimensionless coupling constant, while $\alpha_2$ is a coupling constant of the dimension $[\text{length}]^{2}$. The standard Maxwell Lagrangian $\mathcal{L}_{\text{M}}$ and the quasitopological electromagnetic Lagrangian $\mathcal{L}_{\text{Q}}$ in four dimensions are given by
\begin{eqnarray}
	\mathcal{L}_{\text{M}}&=&-\sqrt{-g}F^2,\\
	\mathcal{L}_{\text{Q}}&=&-\sqrt{-g}\left((F^2)^2-2F^{(4)}\right),\label{LQ}
\end{eqnarray}
where $F^2=-F^\mu{}_\nu F^\nu{}_\mu$, $F^{(4)}=F^\mu{}_\nu F^\nu{}_\rho F^\rho{}_\sigma F^\sigma{}_\mu$, and the Maxwell field strength reads $F_{\mu\nu}=\partial_{\mu}A_{\nu}-\partial_{\nu}A_{\mu}$ with $A_{\mu}$ the vector potential. On the other hand, as a nonlinear electrodynamics, Euler-Heisenberg theory has also attracted significant attentions recently. It was proposed by Euler and Heisenberg~\cite{Heisenberg} and derived directly from quantum electrodynamics to one-loop approximation. And the Schwinger fields are significantly modified accordingly~\cite{Bordin}. However, the quasitopological electromagnetism studied in this work aims at another important property. With the ansatz of the global polarization, such term ensures that the standard Maxwell equation and the energy-momentum tensor are not changed. As a result, this theory is not the most general gauge field theory of the second -order equations.

The energy-momentum tensor of a system with quasitopological electromagnetism reads
\begin{eqnarray}
	T_{\mu\nu}&=&\alpha_1T_{\mu\nu}^{(1)}+\alpha_2T_{\mu\nu}^{(2)}, \nonumber\\
	T_{\mu\nu}^{(1)}&=&2F_{\mu\rho}F_{\nu}{}^{\rho}-\frac{1}{2}F^2g_{\mu\nu}, \nonumber \\ T_{\mu\nu}^{(2)}&=&4F^2F_{\mu\rho}F_{\nu}{}^{\rho}- 8F_{\mu\rho}F^{\rho}{}_{\sigma}F^{\sigma}{}_{\lambda}F^{\lambda}{}_{\nu}-\frac{1}{2}((F^2)^2 -2F^{(4)})g_{\mu\nu}, \label{EqT}
\end{eqnarray}
with the trace given by
\begin{eqnarray}
T^{\mu}_{\;\;\;\mu}=2\alpha_2 \left((F^2)^2-2F^{(4)}\right).
\end{eqnarray}
The Bianchi identity and the Maxwell equation of motion are given, respectively, by
\begin{eqnarray}
	 \nabla_{[\mu}F_{\nu\rho]}=0, \\
	\nabla_{\mu}\widetilde{F}^{\mu\nu}=0,
\end{eqnarray}
where
\begin{eqnarray}
 \widetilde{F}^{\mu\nu}=4\alpha_1F^{\mu\nu}+8\alpha_2(F^2F^{\mu\nu}-2F^{\mu\rho}F^{\sigma}{}_{\rho}F_{\sigma}{}^{\nu}). \label{Max}
\end{eqnarray}
With the ansatz of the global polarization
\begin{eqnarray}
	A_\mu=\xi_\mu \phi(x),
\end{eqnarray}
where $\xi_\mu$ is a constant vector with $\nabla^\mu\xi_\nu=0$, the Maxwell field strength can be written as
\begin{eqnarray}
	F^\mu{}_{\nu}&=&\nabla^\mu A_\nu-\nabla_\nu A^\mu\nonumber\\
	&=&\nabla^\mu(\phi\xi_\nu)-\nabla_\nu(\phi\xi^\mu)\nonumber\\
    &=&\nabla^\mu\phi\xi_\nu+\phi\nabla^\mu\xi_\nu-\nabla_\nu\phi\xi^\mu-\phi\nabla_\nu\xi^\mu\nonumber\\
	&=&\partial^\mu\phi\xi_\nu-\partial_\nu\phi\xi^\mu.
\end{eqnarray}
Furthermore, we can obtain the following equations:
\begin{eqnarray}
	F^{(4)}&=&\frac{1}{2}(F^2)^2\nonumber\\
  &=&2\partial^\mu\phi\partial_\mu\phi\partial^\rho\phi\partial_\rho\phi\xi^\nu\xi_\nu\xi^\sigma\xi_\sigma -4\partial^\mu\phi\partial_\mu\phi\partial^\rho\phi\partial_\sigma\phi\xi^\nu\xi_\nu\xi^\sigma\xi_\rho \nonumber\\
  &\;&+2\partial^\mu\phi\partial_\nu\phi\partial^\rho\phi\partial_\sigma\phi\xi^\nu\xi_\mu\xi^\sigma\xi_\rho,\label{FF1}\\
	F^{\mu\rho}F^\sigma{}_\rho F_\sigma{}^\nu&=&\frac{1}{2}F^2F^{\mu\nu}\nonumber\\
&=&\partial^\alpha\phi\partial_\alpha\phi\xi^\beta\xi_\beta(\partial^\mu\phi\xi^\nu-\partial^\nu\phi\xi^\mu)\nonumber\\
&\;&+\partial^\alpha\phi\xi_\alpha\partial_\beta\xi^\beta(\partial^\nu\phi\xi^\mu-\partial^\mu\phi\xi^\nu),\label{FF2}\\
	F_{\mu\rho}F^\rho{}_\sigma F^\sigma{}_\lambda F^\lambda{}_\nu&=&\frac{1}{2}F^2F_{\mu\rho}F_\nu{}^\rho\nonumber\\
&=&\partial^\alpha\phi\partial_\alpha\phi\xi^\beta\xi_\beta\xi^\rho\xi_\rho\partial_\mu\phi\partial_\nu\phi -\partial^\alpha\phi\xi_\alpha\partial_\beta\phi\xi^\beta\xi^\rho\xi_\rho\partial_\mu\phi\partial_\nu\phi\nonumber\\
&\;&+\partial^\alpha\phi\xi_\alpha\partial_\beta\phi\xi^\beta\xi^\rho\partial_\rho\phi\xi_\mu\partial_\nu\phi -\partial^\alpha\phi\partial_\alpha\phi\xi^\beta\xi_\beta\xi^\rho\partial_\rho\phi\xi_\mu\partial_\nu\phi\nonumber\\
&\;&+\partial^\alpha\phi\xi_\alpha\partial_\beta\phi\xi^\beta\xi^\rho\partial_\rho\phi\partial_\mu\phi\xi_\nu -\partial^\alpha\phi\partial_\alpha\phi\xi^\beta\xi_\beta\xi_\rho\partial^\rho\phi\partial_\mu\phi\xi_\nu\nonumber\\
&\;&+\partial^\alpha\phi\partial_\alpha\phi\xi^\beta\xi_\beta\partial_\rho\phi\partial^\rho\phi\xi_\mu\xi_\nu\nonumber\\
&\;&-\partial^\alpha\phi\xi_\alpha\partial_\beta\phi\xi^\beta\partial^\rho\phi\partial_\rho\phi\xi_\mu\xi_\nu.\label{FF3}
\end{eqnarray}
Thus, it is easy to verify that the quasitopological electromagnetic Lagrangian $\mathcal{L}_{\text{Q}}$ makes no contribution to the Maxwell equation and the energy-momentum tensor by substituting Eqs.~(\ref{FF1}), (\ref{FF2}), and (\ref{FF3}) into $\mathcal{L}_{\text{Q}}$, $T_{\mu\nu}^{(2)}$ and $\widetilde{F}^{\mu\nu}$~\cite{Liu2020}. However, quasitopological electromagnetic can affect the global polarization, which leads to a black hole solution with the coupling $\alpha_2$.

Analogous to thoes in the Einstein-Born-Infeld theory~\cite{LLW2016, AD1982}, dyonic black holes can be constructed in even dimensions. In particular, in four dimensions, the following exact solution for spherically symmetric and static dyonic black holes was obtained~\cite{Liu2020}
\begin{eqnarray}
	ds^2&=&-f dt^2+f^{-1} dr^2+r^2 (d\theta^2+\sin^2\theta d\phi^2),\\
	f(r)&=&-\frac{1}{3}\Lambda r^2+1-\frac{2M}{r}+\frac{\alpha_1 p^2}{r^2}+\frac{q^2}{\alpha_1r^2}~_{2}F_1[\frac{1}{4},1;\frac{5}{4};-\frac{4p^2\alpha_2}{r^4\alpha_1}].
\end{eqnarray}
The Maxwell field for a dyonic particle located at the origin is
\begin{eqnarray}
 F&=&-\phi^{\prime}(r)dt\wedge dr-p\sin\theta d\theta\wedge d\phi,\\
 \phi^\prime(r)&=&-\frac{qr^2}{\alpha_1 r^4+4\alpha_2 p^2}.
\end{eqnarray}
According to the definition of mass~\cite{AD1982,Ashtekar,AshtekarDas} in asymptotically AdS space, we achieve
\begin{eqnarray}
 M_{\rm ADT}=M
\end{eqnarray}
by the Abbott-Deser-Tekin formalism. The parameters $p$ and $q$ are related to the electric charge $Q_{\rm e}$ and magnetic charge $Q_{\rm m}$ of the black hole as follows~\cite{Liu2020}:
\begin{eqnarray}
 Q_{\rm e}=\frac{1}{4\pi}\int\tilde{F}^{0r}=q, \quad
 Q_{\rm m}=\frac{1}{4\pi\alpha_1}\int F=\frac{p}{\alpha_1}. \label{Qm}
\end{eqnarray}
The outer horizon is located at the largest root of $f(r_{\rm h}) =0$. Employing this with the horizon radius $r_{\rm h}$, one can express the black hole temperature and entropy as
\begin{eqnarray}
	T&=&-\frac{q^2r_{\rm h}}{4\pi(4\alpha_2p^2+\alpha_1r_{\rm h}^4)}-\frac{\alpha_1p^2}{4\pi r_{\rm h}^3}-\frac{\Lambda r_{\rm h}}{4\pi}+\frac{1}{4\pi r_{\rm h}},\\
	S&=&\pi r_{\rm h}^2.
\end{eqnarray}
Interpreting the cosmological constant as pressure $P=-\frac{\Lambda}{8\pi}$~\cite{Kastor2009}, the thermodynamic volume, mass, temperature and electric and magnetic potentials of the black hole can be further reexpressed as
\begin{eqnarray}
V&=&\frac{4}{3}\pi r_{\rm h}^3,\\
M&=&\frac{3r_{\rm h}^2\alpha_1+8P\pi r_{\rm h}^4\alpha_1+3Q_{\rm m}^2\alpha_1^4+3Q_{\rm e}^2~_{2}F_1[\frac{1}{4},1;\frac{5}{4};-\frac{4Q_{\rm m}^2\alpha_1\alpha_2}{r_{\rm h}^4}]}{6r_{\rm h}\alpha_1},\\
T&=&\frac{1}{4\pi r_{\rm h}}+2P r_{\rm h}-\frac{Q_{\rm m}^2 \alpha_1^3}{4\pi r_{\rm h}^3}-\frac{Q_{\rm e}^2r_{\rm h}}{4\pi(r_{\rm h}^4\alpha_1+4Q_{\rm m}^2\alpha_1^2\alpha_2)},\label{temp} \\
\Phi_e&=&\frac{Q_{\rm e}~_{2}F_1[\frac{1}{4},1;\frac{5}{4};-\frac{4Q_{\rm m}^2\alpha_1\alpha_2}{r_{\rm h}^4}]}{\alpha_1 r_{\rm h}},\\
\Phi_m&=&-\frac{Q_{\rm e}^2~_{2}F_1[\frac{1}{4},1;\frac{5}{4};-\frac{4Q_{\rm m}^2\alpha_1\alpha_2}{r_{\rm h}^4}]}{4\alpha_1Q_{\rm m}r_{\rm h}}+\frac{Q_{\rm e}^2r_{\rm h}^3}{4Q_{\rm m}(4\alpha_2\alpha_1^2Q_{\rm m}^2+\alpha_1r_{\rm h}^4)}+\frac{\alpha_1^3Q_{\rm m}}{r_{\rm h}}.
\end{eqnarray}
Since parameter $\alpha_2$ has dimensions, we treat it as a thermodynamical variable, and thus it is straightforward to verify the first law of thermodynamics
\begin{eqnarray}
	dM=TdS+\Phi_edQ_{\rm e}+\Phi_mdQ_{\rm m}+\Phi_{\alpha_2}d\alpha_2+VdP,\label{flaw}
\end{eqnarray}
where
\begin{eqnarray}
\Phi_{\alpha_2}=\frac{Q_{\rm e}^2r_{\rm h}^3}{8\alpha_1\alpha_2(r_{\rm h}^4+4\alpha_1\alpha_2Q_{\rm m}^2)}-\frac{Q_{\rm e}^2~_{2}F_1[\frac{1}{4},1;\frac{5}{4};-\frac{4Q_{\rm m}^2\alpha_1\alpha_2}{r_{\rm h}^4}]}{8\alpha_1\alpha_2 r_{\rm h}}
\end{eqnarray}
is the corresponding quantity conjugate to $\alpha_2$. Moreover, the following Smarr relation also holds
\begin{eqnarray}
	M=2TS-2PV+\Phi_eQ_{\rm e}+\Phi_mQ_{\rm m}+2\alpha_2\Phi_{\alpha_2}.
\end{eqnarray}
From the differential form (\ref{flaw}), it is obvious that the black hole mass here plays the role of enthalpy rather than the internal energy of the system. As a result, the Gibbs free energy $G=M-TS$ reads
\begin{eqnarray}
G&=&\frac{1}{12r_{\rm h}\alpha_1(r_{\rm h}^4+4Q_{\rm m}^2\alpha_1\alpha_2)}\Big(3Q_{\rm e}^2r_{\rm h}^4+\alpha_1(3r_{\rm h}^3-8P\pi r_{\rm h}^4+9Q_{\rm m}^2\alpha_1^3)(r_{\rm h}^4+4Q_{\rm m}^2\alpha_1\alpha_2)\nonumber \\
&\;&+6Q_{\rm e}^2(r_{\rm h}^4+4Q_{\rm m}^2\alpha_1\alpha_2)~_{2}F_1[\frac{1}{4},1,\frac{5}{4},-\frac{4Q_{\rm m}^2\alpha_1\alpha_2}{r_{\rm h}^4}]\Big).
\end{eqnarray}
In general, the system always prefers the phase of lowest free energy. Thus, the Gibbs free energy $G$ is an important quantity to study the phase transition. For example, it is well known that the swallowtail behavior of the Gibbs free energy indicates a first-order phase transition of the system. In the following sections, we shall clearly show this for the black hole phase transition.

\section{Phase transitions and phase diagrams}
\label{s3}

In this section, we would like to study the phase transitions and phase diagrams by varying the coupling parameters $\alpha_1$ and $\alpha_2$, respectively.

Reformulating the black hole temperature (\ref{temp}), we obtain the following equation of state for the black hole system
\begin{eqnarray}
 P=\frac{T}{2r_{\rm h}}-\frac{1}{8\pi r_{\rm h}^2}+\frac{Q_{\rm m}^2\alpha_1^3}{8\pi r_{\rm h}^4}+\frac{Q_{\rm e}^2}{8\pi \alpha_1(r_{\rm h}^4+4Q_{\rm m}^2\alpha_1\alpha_2)}.\label{ppres}
\end{eqnarray}
Obviously, the parameters $\alpha_1$ and $\alpha_2$ have potential influence on this equation. As expected, we can define the specific volume $v=2r_{\rm h}$, with which the pressure is cast in the standard form $P=\frac{T}{v}+\mathcal{O}(v)$. On the other hand, since the thermodynamical volume $V\propto r_{\rm h}^3$, the critical point can be determined by
\begin{eqnarray}
 \left(\frac{\partial P}{\partial r_{\rm h}}\right)_T=0,\quad
 \left(\frac{\partial^2 P}{\partial r_{\rm h}^2}\right)_T=0,\label{cond1}
\end{eqnarray}
or alternatively,
\begin{eqnarray}
 \left(\frac{\partial T}{\partial r_{\rm h}}\right)_P=0,\quad
 \left(\frac{\partial^2 T}{\partial r_{\rm h}^2}\right)_P=0.\label{cond2}
\end{eqnarray}
Now let us examine the heat capacity $C_P$ at constant pressure, which measures the local thermodynamical stability of the black holes. A positive or negative value of the heat capacity indicates that the system is locally stable or unstable. After a simple algebraic calculation, we have
\begin{eqnarray}
 C_P=T\left(\frac{\partial S}{\partial T}\right)_P=T\left(\frac{\partial_{r_{\rm h}} S}{\partial_{r_{\rm h}} T}\right)_P\propto \left(\partial_{r_{\rm h}} T\right)_P^{-1},
\end{eqnarray}
where we have used the conditions $T>0$ and $(\partial_{r_{\rm h}} S)_P>0$ for the black hole system. Therefore, in the $T$-$r_{\rm h}$ plane, the black hole branch with a positive or negative slope is thermodynamically stable or unstable. We shall distinguish them by solid or dashed curves in the following.

Here we note that the four-dimensional charged RN-AdS black hole will be obtained by setting $\alpha_1=1$ and $\alpha_2=p=0$. The equation of state (\ref{ppres}) reduces to~\cite{Kubiznak2012}
\begin{eqnarray}
 P=\frac{T}{2r_{\rm h}}-\frac{1}{8\pi r_{\rm h}^2}+\frac{Q_{\rm e}^2}{8\pi r_{\rm h}^4}.
\end{eqnarray}
Employing this with the condition (\ref{cond1}), the critical point will be obtained as
\begin{eqnarray}
 P_{\rm c}=\frac{1}{96\pi Q_{\rm e}^2}, \quad
 T_{\rm c}=\frac{\sqrt{6}}{18\pi Q_{\rm e}},\quad
 r_{\rm hc}=\sqrt{6}Q_{\rm e}.
\end{eqnarray}
As shown in Ref.~\cite{Kubiznak2012}, only one small-large black hole phase transition exists, which is reminiscent of the liquid/gas phase transition of the vdW fluid.

Next, we shall focus on the black hole phase transition by setting the electric charge $Q_{\rm e}=5$ and the magnetic charge $Q_{\rm m}=2.5$. Then different characteristic patterns of the phase transitions and phase diagrams will be displayed by varying $\alpha_1$ and $\alpha_2$. Here we briefly describe our process: (i) By making use of Eq. (\ref{cond1}) or (\ref{cond2}), we obtain the critical points first. (ii) Then we examine the the local stability for these black hole branches along each isobaric curve in the $T$-$r_{\rm h}$ plane. These black hole branches with positive and negative slopes marked with solid and dashed curves are thermodynamically stable and unstable, respectively. The extremal points of the temperature correspond to the nonsmooth points in the $G$-$T$ plane. (iii) Next, the phase transition points for each given pressure will be obtained by analyzing the swallowtail behavior of the Gibbs free energy. Here one should be very careful to ensure that only the swallowtail behaviors constructed by two stable black hole branches indicate the phase transitions. (iv) Varying the pressure freely, we shall obtain the coexistence curves of the phase transition, with which the phase diagrams can also be explicitly shown in different parameter spaces. And in this paper we are concerned mainly with the pressure-temperature diagram. (v) Finally, critical exponents will be calculated for the critical points that we consider.

\subsection{Phase transitions by varying $\alpha_2$}

In this subsection, we would like to study the phase transition and phase diagram for the black hole by fixing $\alpha_1=1$ while varying $\alpha_2$. In order to show the characteristic behaviors of the phase transition, we take $\alpha_2$=15, 45, and 75, respectively.

\subsubsection{$\alpha_2=15$}

\begin{figure}
	\center{\subfigure[]{\label{Fig1a}
			\includegraphics[width=5cm]{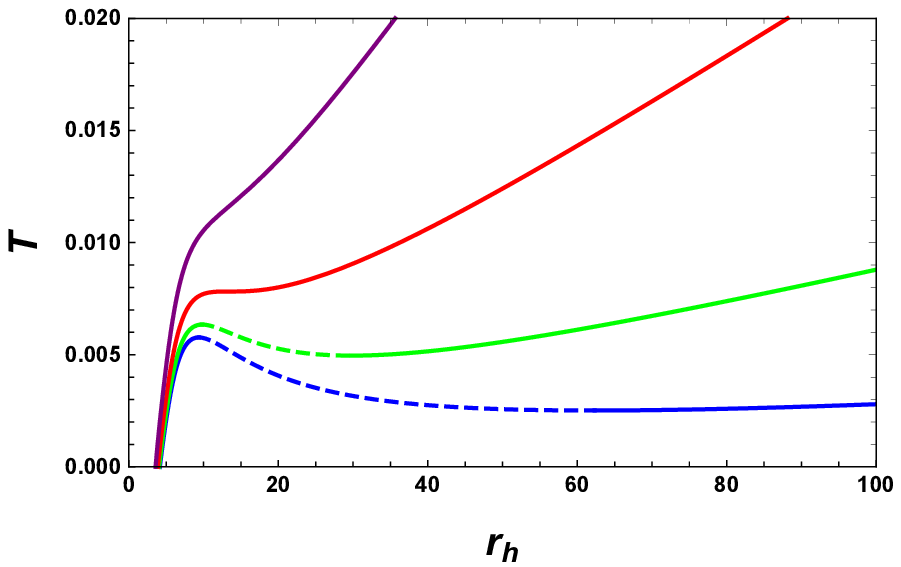}}
		\subfigure[]{\label{Fig1b}
			\includegraphics[width=5cm]{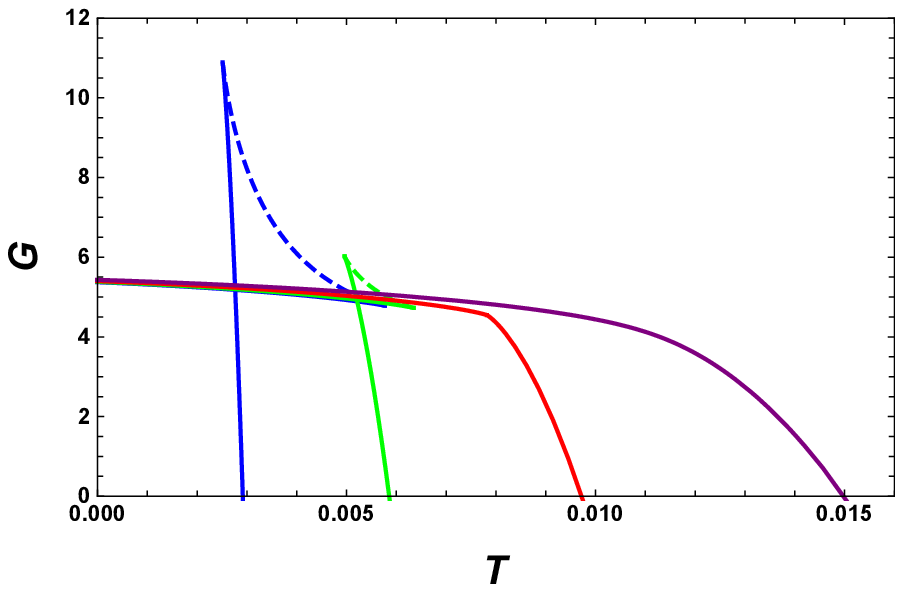}}
		\subfigure[]{\label{Fig1c}
			\includegraphics[width=5cm]{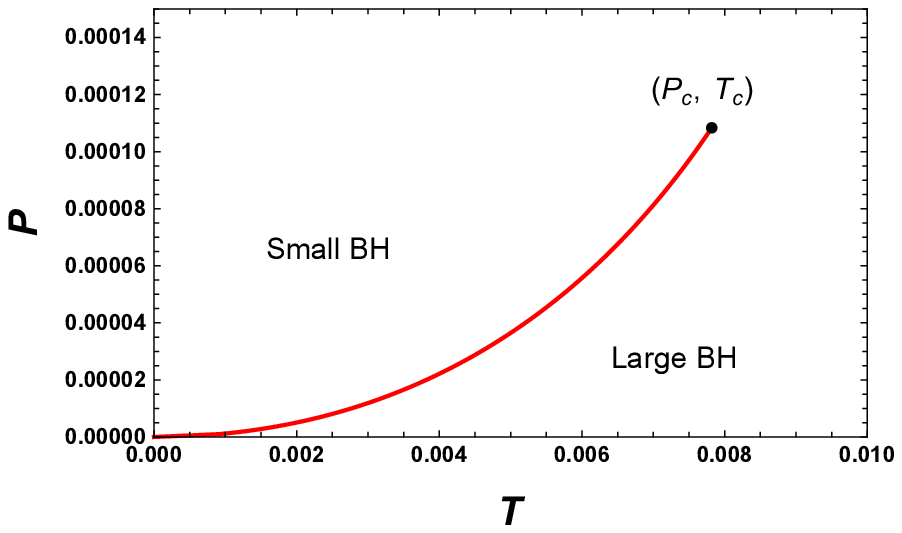}}}
\caption{(a) $T$ vs $r_{\rm h}$ and (b) $G$ vs $T$. The pressure $P=0.00001$ (blue curves), $0.00004$ (green curves), $0.00010836$ (red curves) and $0.00025$ (purple curves), displayed from bottom to top in (a) and from left to right in (b). The unstable and stable branches are described using dashed and solid curves, respectively. (c) $P$-$T$ phase diagram. Coupling parameters $\alpha_1=1$ and coupling parameter $\alpha_2=15$.}\label{Fig1}
\end{figure}

Adopting the condition (\ref{cond1}), we find there is one critical point at
\begin{equation}
 T_{\rm c}=0.00781870,\quad P_{\rm c}=0.00010836.
\end{equation}
The behaviors of the temperature $T$ and the Gibbs free energy $G$ are plotted in Figs. \ref{Fig1a} and \ref{Fig1b}. In Fig. \ref{Fig1a}, one can find that for each isobaric curve with pressure $P<P_{\rm c}$, there are two extremal points which divide the isobaric curve into three branches: the stable small and large black hole branches (described with the solid curves), and the unstable intermediate black hole branch (described with the dashed curves). These stable and unstable black hole branches are shown accordingly in \ref{Fig1b}. Meanwhile, the two extremal points at each isobaric curve also correspond to the nonsmooth points in the $G$-$T$ diagram. When $P<P_{\rm c}$, the swallowtail behaviors are present. Since they are constructed using the stable small and large black hole branches, this indeed indicates the phase transitions. For each isobaric curve, we can find that, with an increase in temperature, the system prefers a small black hole phase first, then turns to a large black hole phase after the intersection point is approached.

With an increase in the pressure $P$, these two extremal points of temperature along the isobaric curve get closer, and they coincide at the critical point at which the critical pressure is approached. The swallowtail behavior of the Gibbs free energy disappears at the critical case. For $P>P_{\rm c}$, the temperature monotonically increases with the radius of the black hole horizon, and the Gibbs free energy monotonically decreases with the temperature, so no phase transitions can be observed.

Finally, we illustrate the coexistence curve of the small and large black hole phases in the $P$-$T$ diagram in Fig. \ref{Fig1c}. Obviously, the pressure increases monotonically with the temperature, and terminates at a critical point, at which the black hole undergoes a second-order phase transition. This is a typical small-large black hole phase transition of the vdW-like type.

\subsubsection{$\alpha_2=45$}

\begin{figure}
	\center{\subfigure[]{\label{Fig2a}
			\includegraphics[width=5cm]{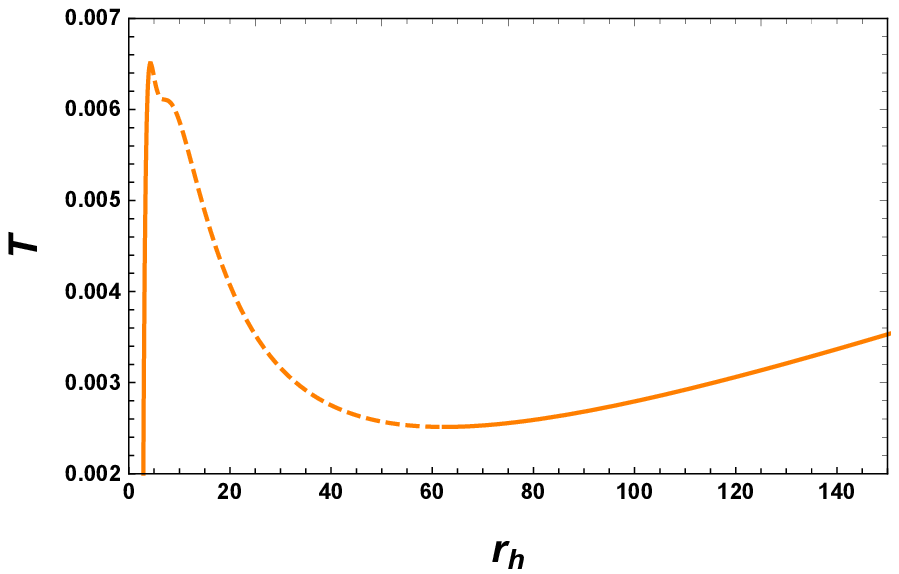}}
		\subfigure[]{\label{Fig2b}
			\includegraphics[width=5cm]{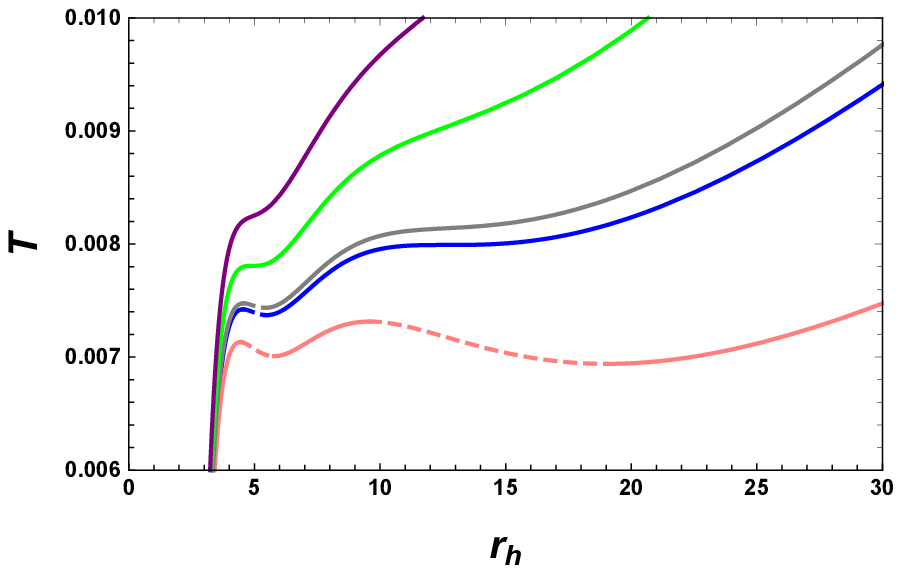}}
		\subfigure[]{\label{Fig2c}
			\includegraphics[width=5cm]{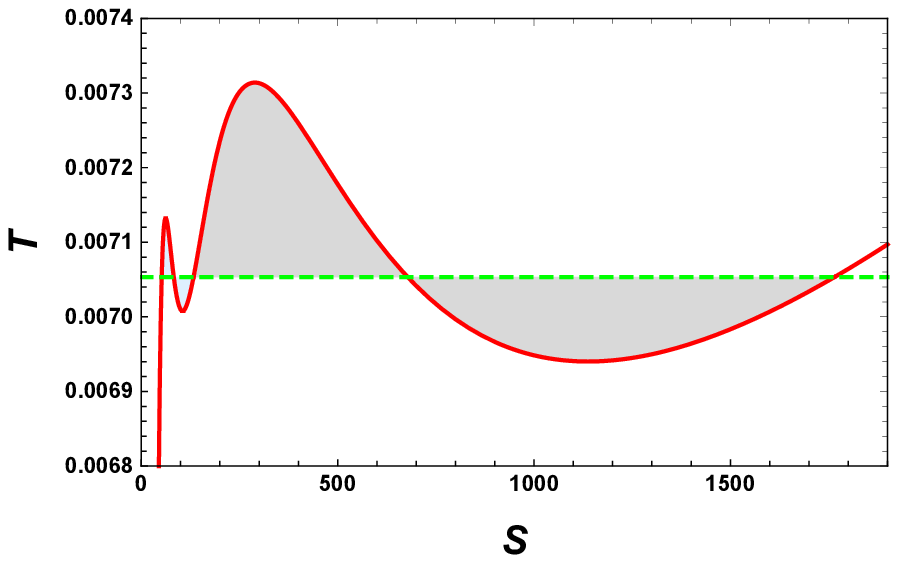}}}
\caption{(a) Behavior of $T$ with respect to $r_{\rm h}$ for $P=0.00001<P_{\rm c1}$ (orange curve). (b) Behavior of $T$ with respect to $r_{\rm h}$ (pressure larger than $P=P_{\rm c1}=0.00001478$) for $0.00008189$ (pink curve), $0.00011414$ (blue curve), $0.00012$ (gray curve), $0.00015547$ (green curve) and $0.0002$ (purple curve) displayed from bottom to top. (c) $T$-$S$ diagram for ``double" Maxwell equal area laws with a pressure $P=P_{\rm t}=0.00008189$. The horizontal line has a temperature $T=T_{\rm t}=0.00705331$, and parameters $\alpha_1=1$, $\alpha_2=45$. The unstable and stable branches are indicated by dashed and solid curves, respectively.}\label{Fig2}
\end{figure}

Unlike the previous case, when the parameter $\alpha_2=45$, we observe three critical points
\begin{eqnarray}
 T_{\rm c1}=0.00617553, \quad P_{\rm c1}&=&0.00001478,\\
 T_{\rm c2}=0.00799143, \quad P_{\rm c2}&=&0.00011414,\\
 T_{\rm c3}=0.00780718, \quad P_{\rm c3}&=&0.00015547.
\end{eqnarray}
At first, let us focus on the behavior of the temperature. For different values of the pressure, we plot the temperature as a function of the horizon radius $r_{\rm h}$ in Figs.~\ref{Fig2a} and ~\ref{Fig2b}. For $P<P_{\rm c1}$, which is shown in Fig.~\ref{Fig2a}, there are two extremal points along each isobaric curve. And they divide the isobaric curve into three parts, respectively, corresponding to the stable small black hole branch, the unstable intermediate black hole branch, and the stable large black hole branch. Further increasing the pressure such that $P_{\rm c1}<P<P_{\rm c2}$, an extra new phase and two extremal points emerge, as seen in Fig.~\ref{Fig2b}. Considering their local stability and size, we can name these five branches the stable small black hole branch, the unstable small black hole branch, the stable intermediate black hole branch, the unstable large black hole branch, and the stable large black hole branch. As a result, the novel behavior of the isobaric curve allows us to construct two pairs of equal areas. Taking $P$=$P_{\rm t}=0.00008189$ as an example, we show these two pair areas in Fig.~\ref{Fig2c}. Each pair area is constructed using the Maxwell equal area law, while the values of these two pair areas are not required to be the same. Moreover, one can find that these two pair areas share the same temperature $T_{\rm t}=0.00705331$, which indicates that there are two black hole phase transitions at the same temperature and pressure. This feature actually implies a triple point, where the small, intermediate, and large black holes coexist. On the other hand, it is also worth pointing out that the Maxwell equal area law holds in the $T$-$S$ plane but dose not in the $T$-$r_{\rm h}$ plane.

At $P=P_{\rm c2}$, the system undergoes an intermediate-large black hole phase transition of second order. Above this value, i.e., $P_{\rm c2}<P<P_{\rm c3}$, two black hole branches disappear and only three are left: the the stable small and intermediate black hole branches and the unstable intermediate black hole branch. This result suggests that there is, at most, only the stable small-intermediate black hole phase transition. This first-order phase transition turns to a second-order one when the pressure tends to $P=P_{\rm c3}$, beyond which the temperature becomes a monotonic function of $r_{\rm h}$. Thus, only one black hole branch is left, and no phase transition  exists.

\begin{figure}
	\center{\subfigure[]{\label{Fig3a}
			\includegraphics[width=7cm]{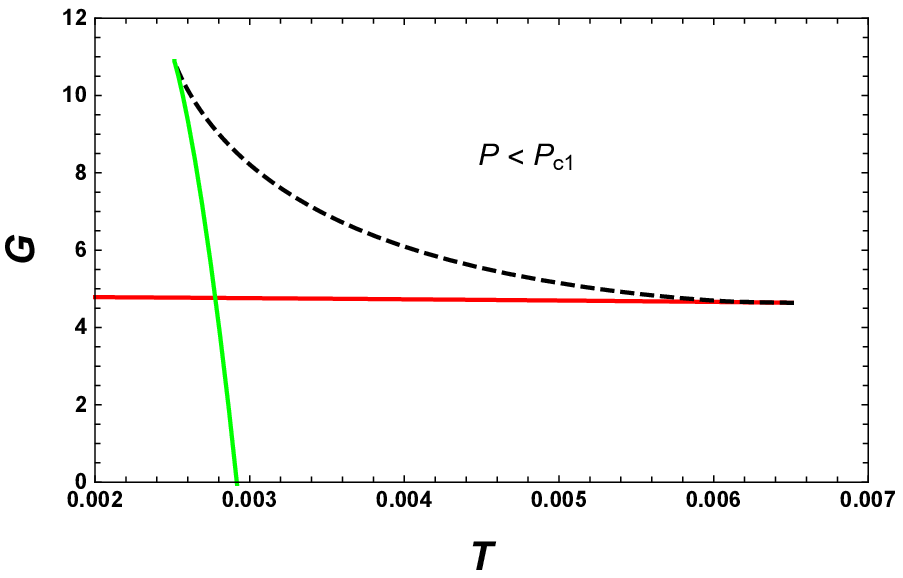}}
		\subfigure[]{\label{Fig3b}
			\includegraphics[width=7cm]{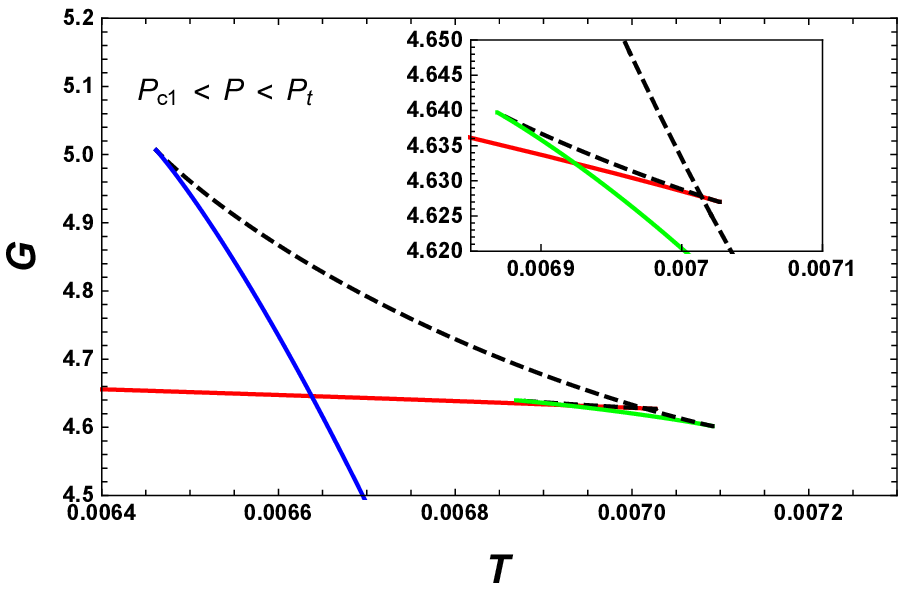}}
		\subfigure[]{\label{Fig3c}
			\includegraphics[width=7cm]{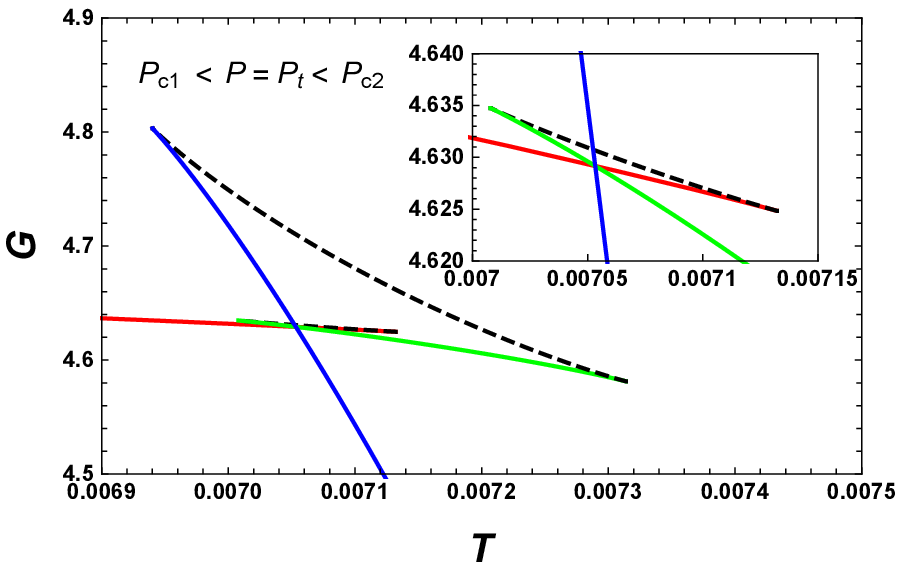}}
		\subfigure[]{\label{Fig3d}
			\includegraphics[width=7cm]{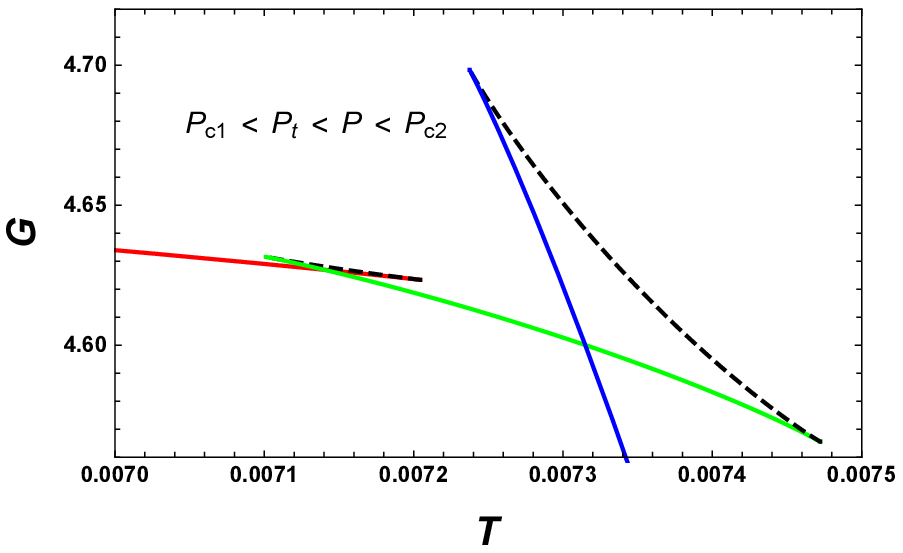}}
		\subfigure[]{\label{Fig3e}
			\includegraphics[width=7cm]{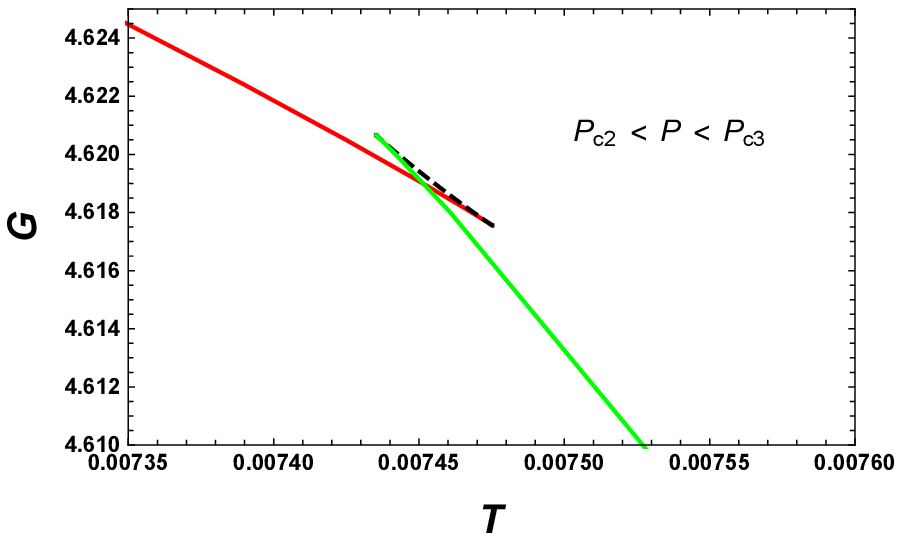}}
		\subfigure[]{\label{Fig3f}
			\includegraphics[width=7cm]{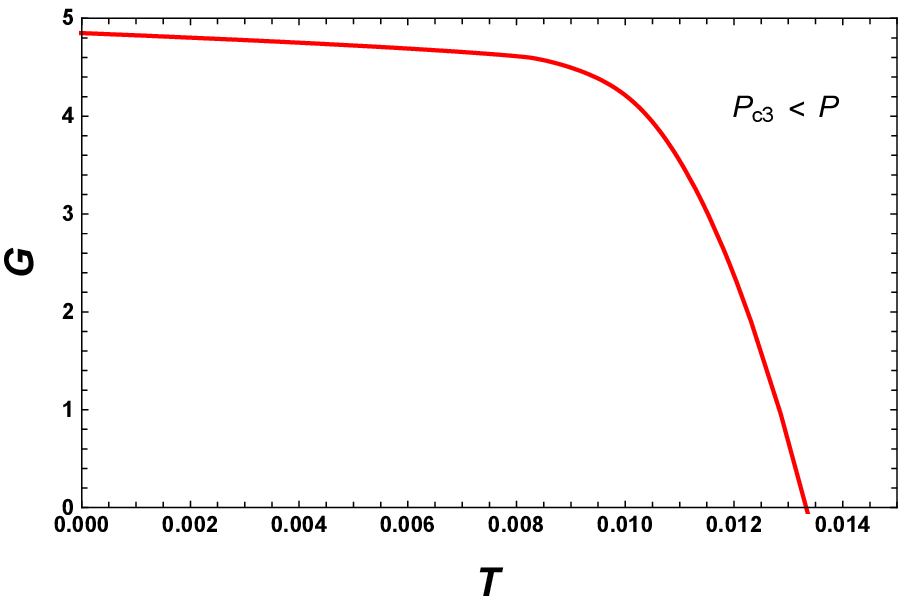}}}
	\caption{Behaviors of the Gibbs free energy $G$ with respect to the temperature $T$ with $\alpha_1=1$ and $\alpha_2=45$. The red, green and blue curves represent small, intermediate and large black holes, respectively. (a) $P=0.00001$. (b) $P=0.00007$. (c) $P=P_{\rm t}=0.00008189$. (d) $P=0.00009$. (e) $P=0.00012$. (f) $P=0.0002$.}\label{Fig3}
\end{figure}

Next, we turn to the Gibbs free energy $G$. Its characteristic behaviors are shown in Fig. \ref{Fig3}. It is worth pointing out that the nonsmooth points of $G$ correspond exactly to the extremal points shown in Figs. \ref{Fig2a} and \ref{Fig2b}. Similarly, the stable and unstable black hole branches are plotted with the solid and dashed curves, respectively. For a small pressure with $P=0.00001<P_{\rm c1}$, the Gibbs free energy $G$ is plotted as a function of the temperature in Fig. \ref{Fig3a}. Obviously, there is a swallowtail behavior, thus indicating the existence of the small-large black hole phase transition, which is similar to the liquid/gas phase transition of vdW fluid. Taking $P_{\rm c1}<P=0.00007<P_{\rm t}$, we exhibit the Gibbs free energy in Fig. \ref{Fig3b}. Interestingly, two swallowtail behaviors can be observed. At first glance, it appears that there may be two phase transitions. However, we note that one of them has a higher free energy. Thus it will be suppressed by the black hole branch with lower free energy and not participate in the black hole phase transition. As a result, there is still only one small-large black hole phase transition. When the pressure is further increased, this swallowtail behavior is shifted to the left. When $P=P_{\rm t}$, as shown in Fig. \ref{Fig3c}, these two intersection points of the swallowtail behaviors coincide exactly. Meanwhile, the three stable black hole branches intersect at that point. This result strongly indicates that the stable small, intermediate and large black holes can coexist at the point namely, a triple point $(T_{\rm t}, P_{\rm t})$. This is also the most distinctive feature of the dyonic AdS black holes.

In Fig. \ref{Fig3d}, we take $P_{\rm t}<P=0.00009<P_{\rm c2}$. Two swallowtail behaviors are also observed. However, unlike in the case shown in Fig. \ref{Fig3b}, both of them can participate in phase transitions. Namely, there will be a first-order small-intermediate black hole phase transition and an intermediate-large black hole phase transition at the same pressure but at different temperatures.

Moreover, the intermediate-large black hole phase transition tends to disappear at $P=P_{\rm c2}$. When the pressure is further increased such that $P_{\rm c2}<P=0.00012<P_{\rm c3}$, one of the swallowtail behaviors disappears; see Fig. \ref{Fig3e}. Thus for each pressure, there is only one first-order phase transition. Detailed study reveals that this phase transition extends to the third critical point at $P=P_{\rm c3}$. While beyond the third critical point, i.e., $P=0.0002>P_{\rm c3}$ displayed in Fig. \ref{Fig3f}, the free energy is a monotonic function of the temperature. Thus no phase transition exists anymore.

\begin{figure}[htb]
	\centering
	\includegraphics[width=8cm]{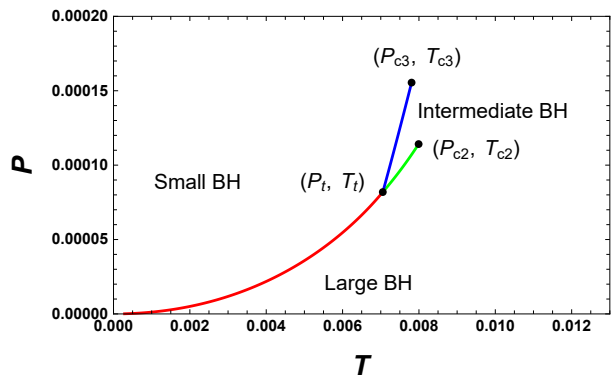}
	\caption{Phase diagram for a dyonic AdS black hole (BH) with $\alpha_1=1$ and $\alpha_2=45$. The characteristic triple point is present.}\label{fig4}
\end{figure}

The corresponding phase diagram is displayed in the $P$-$T$ diagram in Fig.~\ref{fig4}. It exhibits the characteristic triple point feature. Below $P_{\rm t}$, the black hole system undergoes a first-order small-large black hole phase transition. And at $P$=$P_{\rm t}$, the stable small, intermediate and large black hole phases coexist. Furthermore, we can also find that the small-intermediate black hole phase transition emerges at $P_{\rm t}$ and terminates at $P_{\rm c3}$, while the intermediate-large black hole phase transition starts at $P_{\rm t}$ and ends at $P_{\rm c2}$.

\subsubsection{$\alpha_2=75$}

\begin{figure}
\center{\subfigure[]{\label{Fig5a}
			\includegraphics[width=5cm]{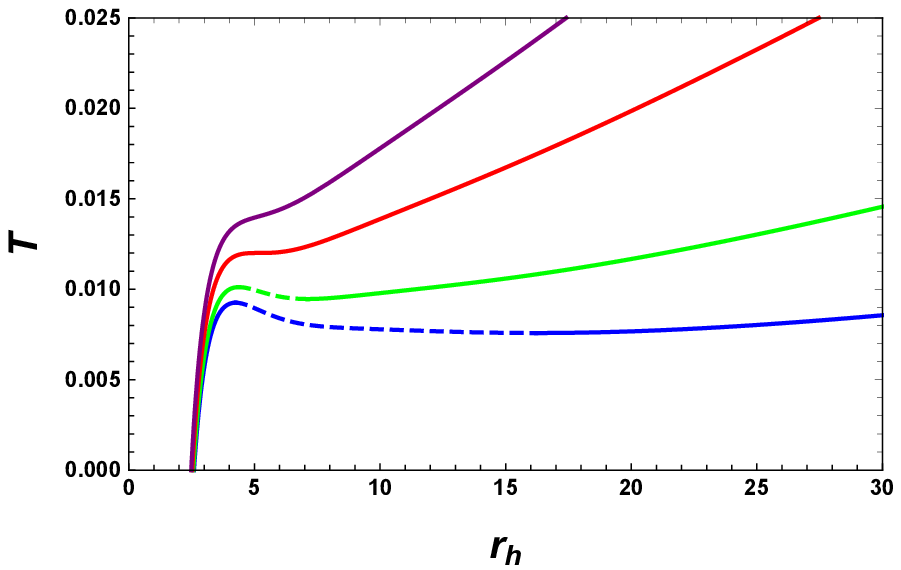}}
		\subfigure[]{\label{Fig5b}
			\includegraphics[width=5cm]{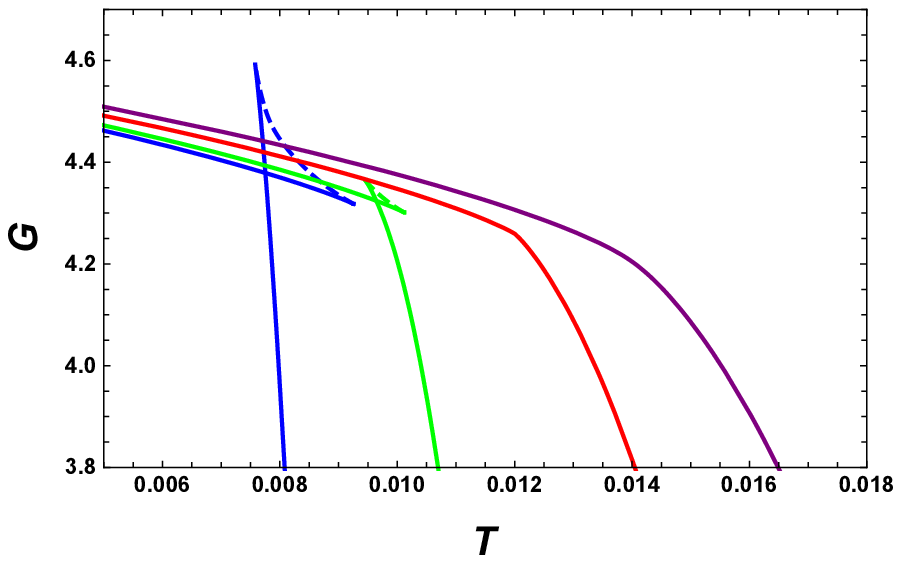}}
		\subfigure[]{\label{Fig5c}
			\includegraphics[width=5cm]{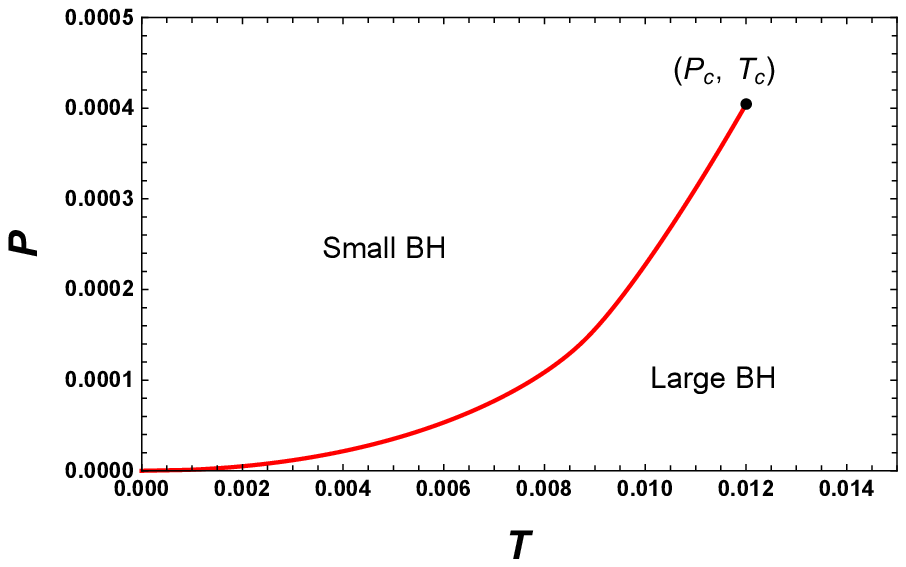}}}
\caption{(a) $T$ vs $r_{\rm h}$. (b) $G$ vs $T$. Pressures $P=0.0001$ (blue curve), $0.0002$ (green curve), $0.00040458$ (red curve) and $0.0006$ (purple curve), displayed from bottom to top in (a) and from left to right in (b). (c) $P$-$T$ phase diagram. The parameters $\alpha_1=1$ and $\alpha_2=75$.}\label{Fig5}
\end{figure}

If we increase the parameter such that $\alpha_2=75$, the number of critical points returns to 1. The value of the critical point is
\begin{eqnarray}
 T_{\rm c}=0.01200427, \quad P_{\rm c}=0.00040458.
\end{eqnarray}
The behaviors of the temperature and the Gibbs free energy are exhibited in Figs. \ref{Fig5a} and \ref{Fig5b} for different pressure values. When the pressure is below its critical value, there are the nonmonotonic behavior of the temperature and the swallowtail behavior of the Gibbs free energy, which indicates the existence of the standard small-large black hole phase transition. The coexistence curve is also shown in Fig. \ref{Fig5c}, which starts at the origin and ends at the critical point marked with a black dot. Above the curve is the small black hole region and below it is the large black hole region as expected.

In summary, besides the small-large black hole phase transition, we also observe a triple point for the dyonic AdS black holes. This feature is quite different from that of the four-dimensional charged RN-AdS black hole, where only the small-large black hole phase transition has been discovered~\cite{Kubiznak2012}.

\subsection{Phase transition when varying $\alpha_1$}

In the previous subsection, we showed that there is a small-large black hole phase transition and triple point when $\alpha_1$ is fixed but $\alpha_2$ is varied freely. Since $\alpha_1$ is also an important parameter of the dyonic AdS black hole, in this subsection we shall vary the parameter $\alpha_1$ instead while keeping $\alpha_2$=50. One can also see that novel interesting phase transitions will be disclosed.

\subsubsection{$\alpha_1=0.4$}

We first take $\alpha_1=0.4$. After solving Eq.~(\ref{cond1}), we have two critical points
\begin{eqnarray}
 T_{\rm c1}&=&0.00548121, \quad P_{\rm c1}=0.00005316, \\
 T_{\rm c2}&=&0.06782108, \quad P_{\rm c2}=0.01299062.
\end{eqnarray}
Since the values of these two critical points are quite different, they are located far away from each other in the phase diagram.

\begin{figure}
	\center{\subfigure[]{\label{Fig6a}
			\includegraphics[width=7cm]{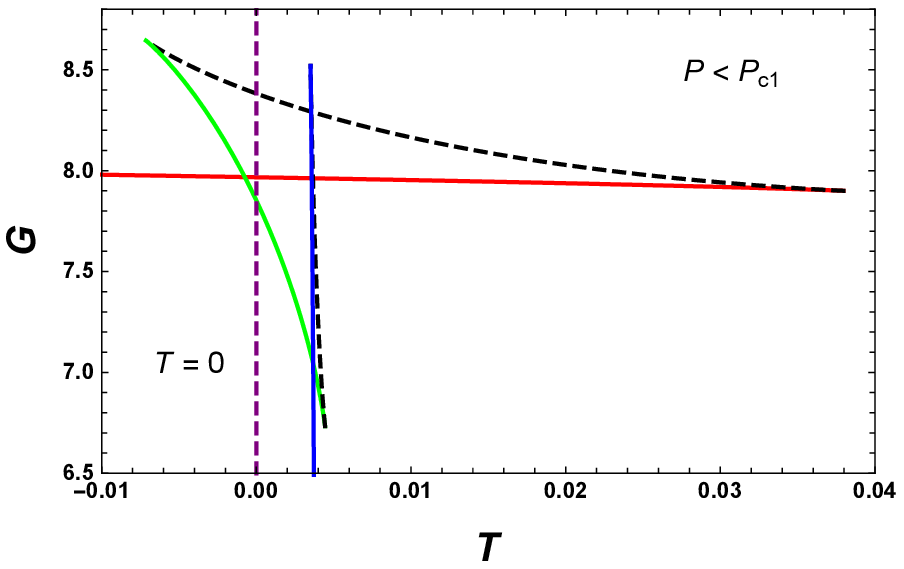}}
		\subfigure[]{\label{Fig6b}
			\includegraphics[width=7cm]{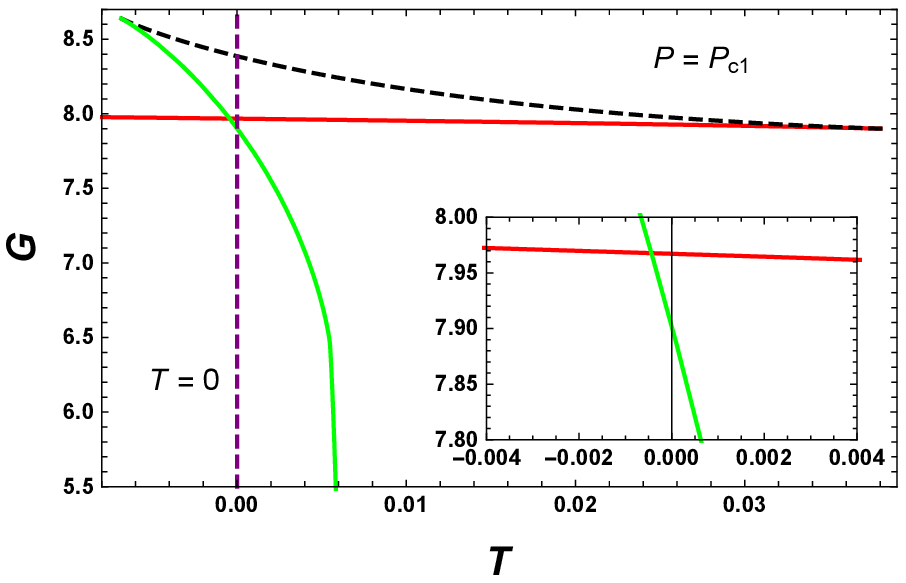}}
		\subfigure[]{\label{Fig6c}
			\includegraphics[width=7cm]{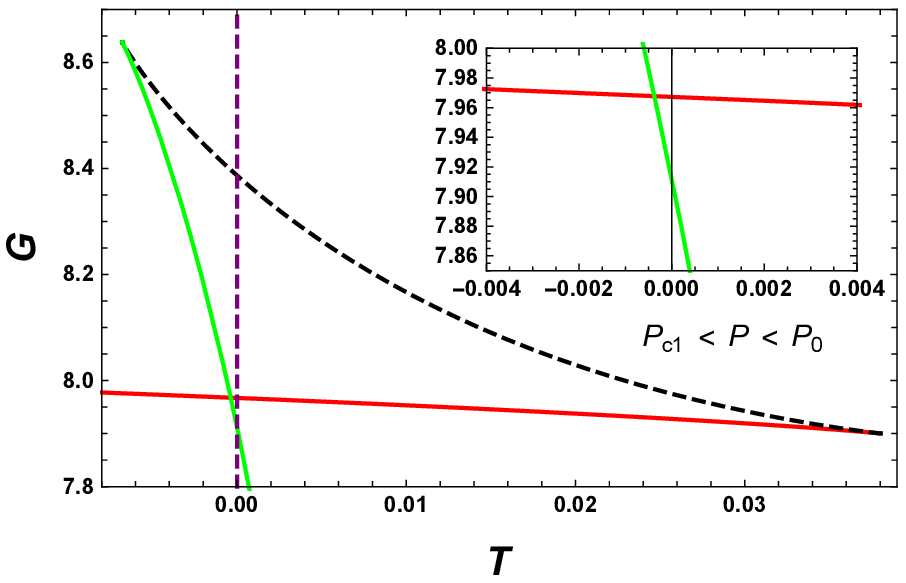}}
		\subfigure[]{\label{Fig6d}
			\includegraphics[width=7cm]{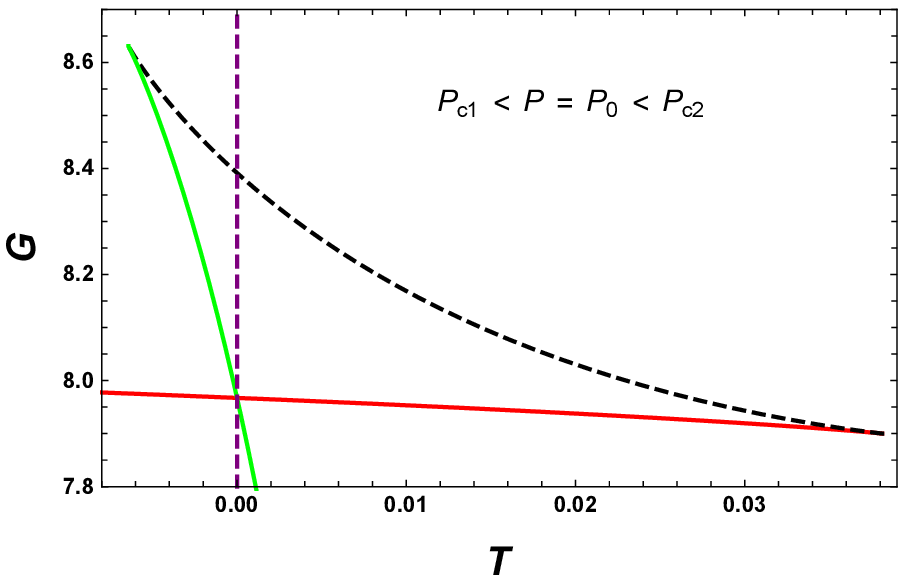}}
		\subfigure[]{\label{Fig6e}
			\includegraphics[width=7cm]{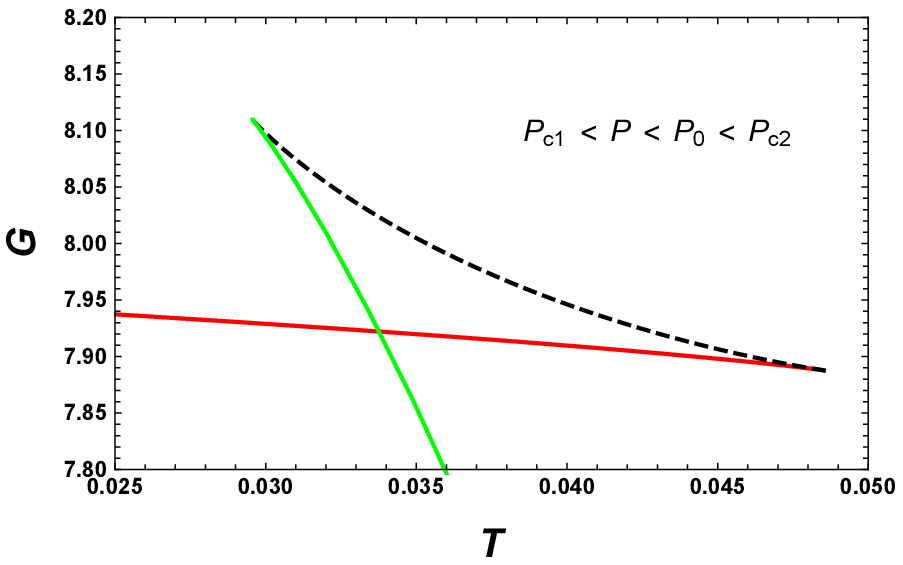}}
		\subfigure[]{\label{Fig6f}
			\includegraphics[width=7cm]{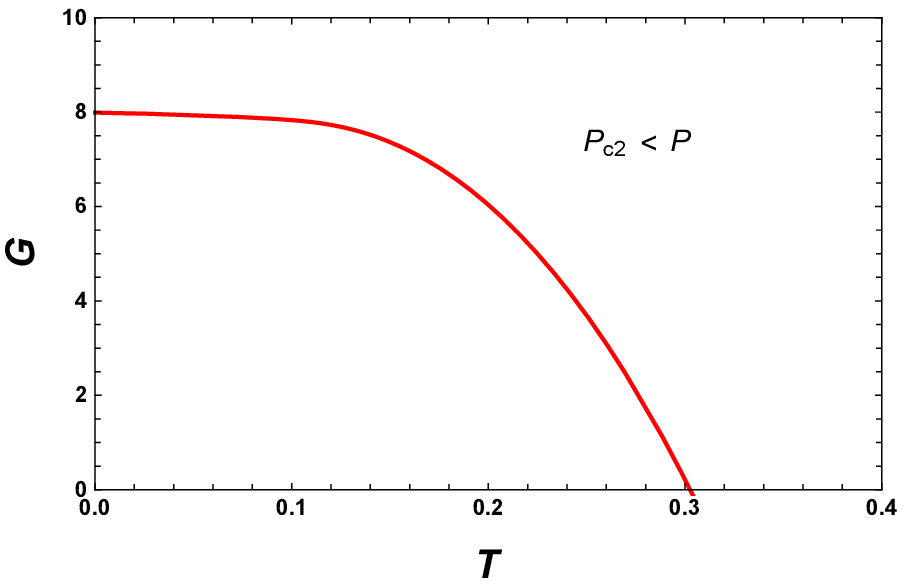}}}
	\caption{Behaviors of the Gibbs free energy $G$ with respect to the temperature $T$ with $\alpha_1=0.4$ and $\alpha_2=50$. The red, green and blue curves represent small, intermediate and large black holes, respectively. The purple dashed vertical lines indicate vanishing temperature. (a) $P=0.00002$. (b) $P=P_{\rm c1}=0.00005316$. (c) $P=0.00006$. (d) $P=P_0=0.00010129$. (e) $P=0.005$. (f) $P=0.03$.}\label{Fig6}
\end{figure}

The interesting behaviors of Gibbs free energy $G$ are displayed in Fig. \ref{Fig6}. When the pressure $P<P_{\rm c1}$, as shown in Fig. \ref{Fig6a}, two swallowtail behaviors are present. However, one of them is located in the negative temperature region, and thus it is unphysical. As a result, only the other swallowtail behavior is available, which indicates an intermediate-large black hole phase transition of the first order. In Fig.~\ref{Fig6b}, we plot $G$ as a function of the temperature $T$ at the critical pressure $P=P_{\rm c1}$. The swallowtail behavior with positive temperature tends to disappear. For this case, only a second-order phase transition exists. Increasing the pressure such that $P_{\rm c1}<P<P_0=0.00010129$, we can see in Fig. \ref{Fig6c} that the intersection point of the swallowtail behavior is still located at the negative temperature, and thus such behavior does not indicate a phase transition. With detailed study, we find that the intersection point is shifted to the right. Exactly at $P=P_0$, as shown in Fig. \ref{Fig6d},  the intersection point has a vanishing temperature. If we continue to slightly increase the pressure [see Fig. \ref{Fig6e}], the intersection point has a positive temperature, thereby indicating a first-order phase transition. At last, by setting $P>P_{\rm c2}$, the $G$ described in Fig. \ref{Fig6f} is only a monotonically decreasing function of temperature. Thus, the system has only one black hole branch and no phase transition is allowed.

\begin{figure}
	\center{\subfigure[]{\label{fig7a}
		\includegraphics[width=7cm]{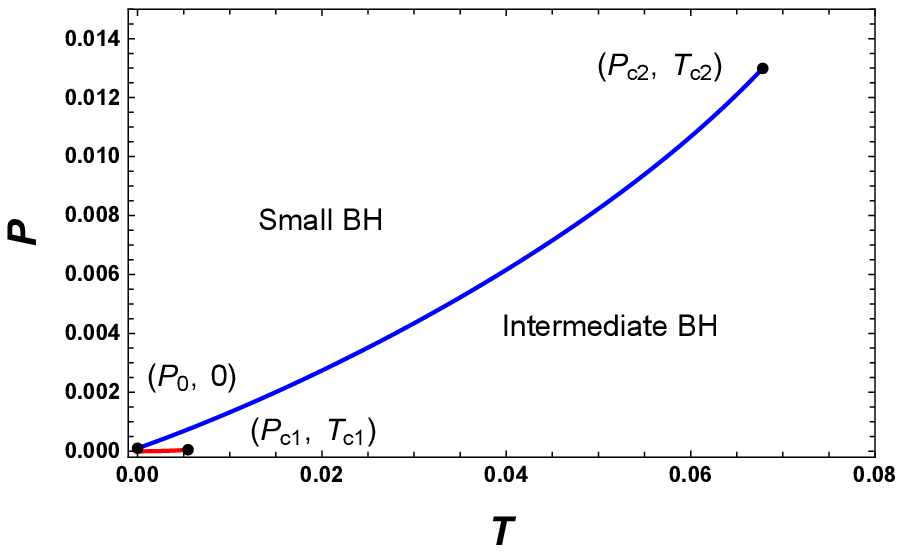}}
	\subfigure[]{\label{fig7b}
		\includegraphics[width=7.5cm]{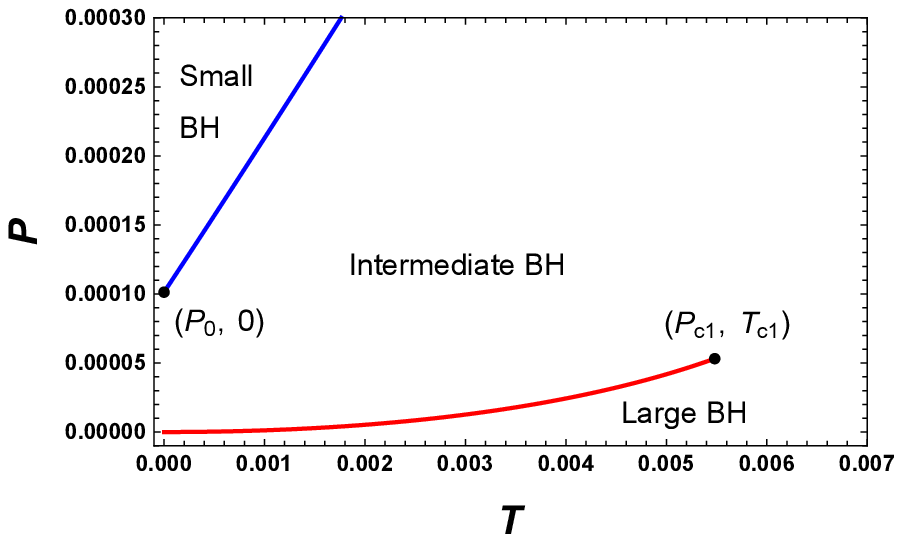}}}
	\caption{Phase diagram for a dyonic AdS black hole with $\alpha_1=0.4$ and $\alpha_2=50$. (a) Entire phase diagram. (b) Magnification of (a) near ($T_{\rm c1}$, $P_{\rm c1}$).}\label{fig7}
\end{figure}

Considering these different behaviors of the Gibbs free energy, we calculate the phase transition for each temperature. Finally, we clearly exhibit the phase diagram of the dyonic AdS black hole with $\alpha_1=0.4$ and $\alpha_2=50$ in the $P$-$T$ diagram in Fig. \ref{fig7}. In the figure, it is easy to see that this phase diagram is significantly different from that of the vdW fluid. There are two separate coexistence curves. One of them starts at the origin and ends at $(T_{\rm c1}, P_{\rm c1})$, while the other extends from $(0, P_0)$ and terminates at the second critical point $(T_{\rm c2}, P_{\rm c2})$. This novel phase structure has not been observed for the charged RN-AdS black hole~\cite{Kubiznak2012}, and can be treated as a new feature of dyonic AdS black holes with quasitopological electromagnetism.

\subsubsection{$\alpha_1=1$}

\begin{figure}
	\center{\subfigure[]{\label{Fig8a}
			\includegraphics[width=7cm]{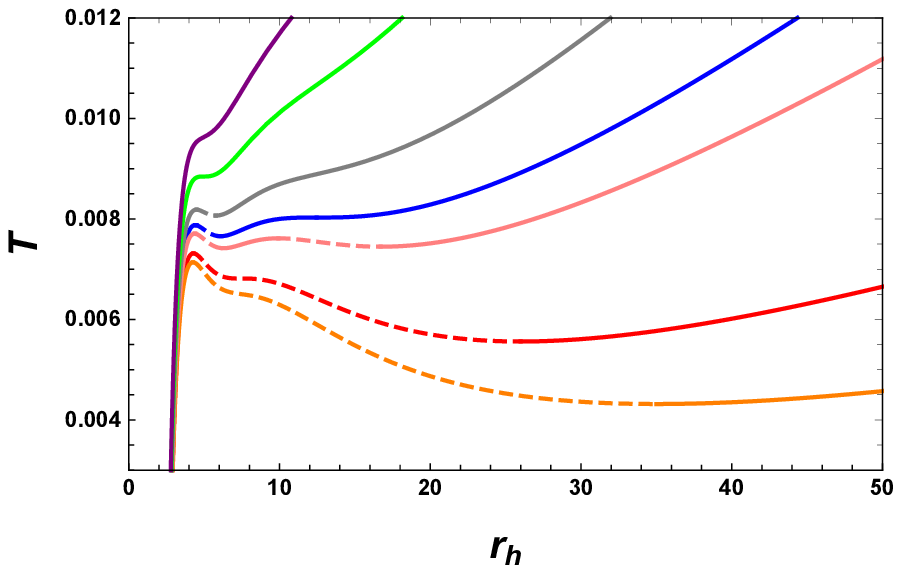}}
		\subfigure[]{\label{Fig8b}
			\includegraphics[width=7cm]{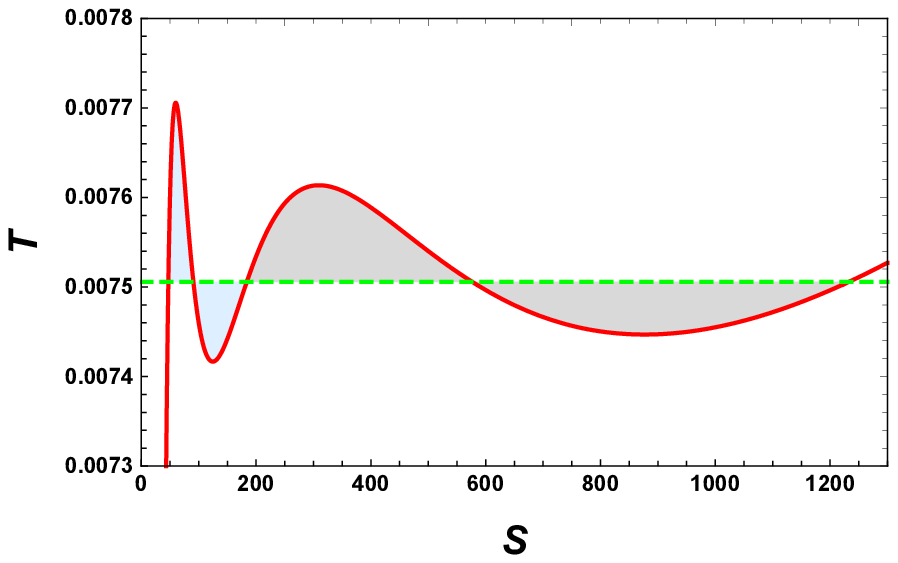}}}
	\caption{(a) Behavior of $T$ with respect to $r_{\rm h}$ for $P=0.00003$ (orange curve), $P=0.00005078$ (Red curve), $P=0.00009608$ (pink curve), $P=0.00011537$ (blue curve), $P=0.00015$ (gray curve), $P=0.00022122$ (green curve) and $P=0.0003$ (purple curve), displayed from bottom to top. (b) The Maxwell equal area laws constructed at the isobaric curve with pressure $P=P_{\rm t}=0.00009608$ in the $T$-$S$ diagram. The horizontal line has a temperature $T=T_{\rm t}=0.00750572$. The parameters $\alpha_1=1$ and $\alpha_2=50$.}\label{Fig8}
\end{figure}

\begin{figure}
	\center{\subfigure[]{
			\includegraphics[width=7cm]{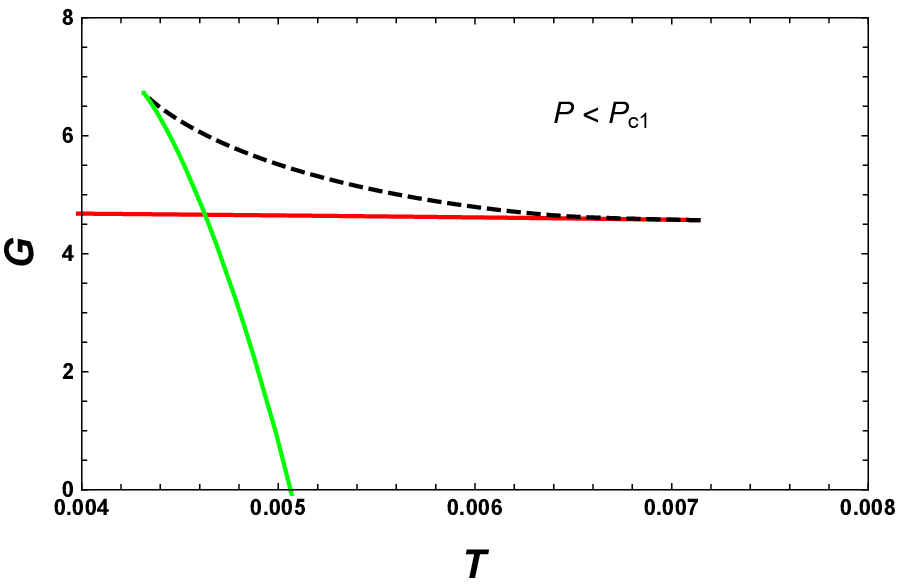}}
		\subfigure[]{
			\includegraphics[width=7cm]{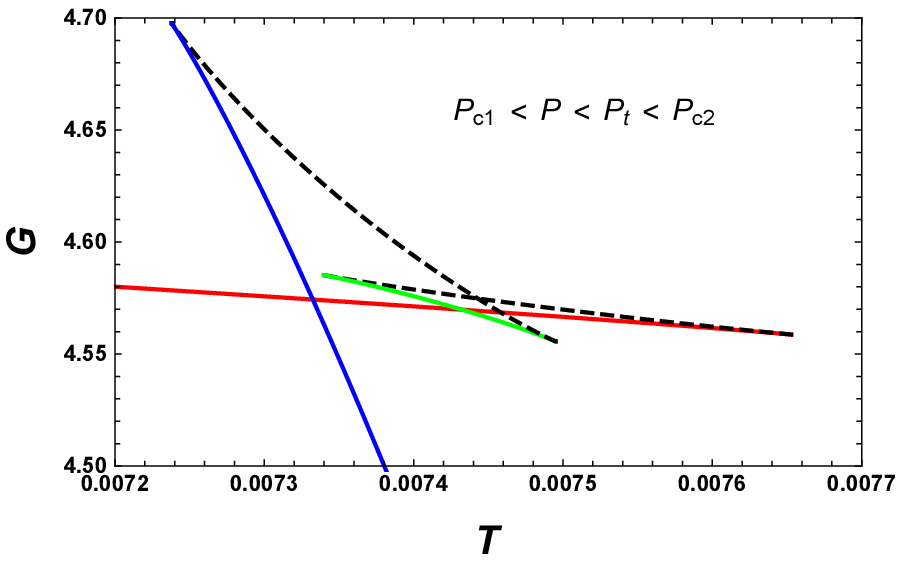}}
		\subfigure[]{
			\includegraphics[width=7cm]{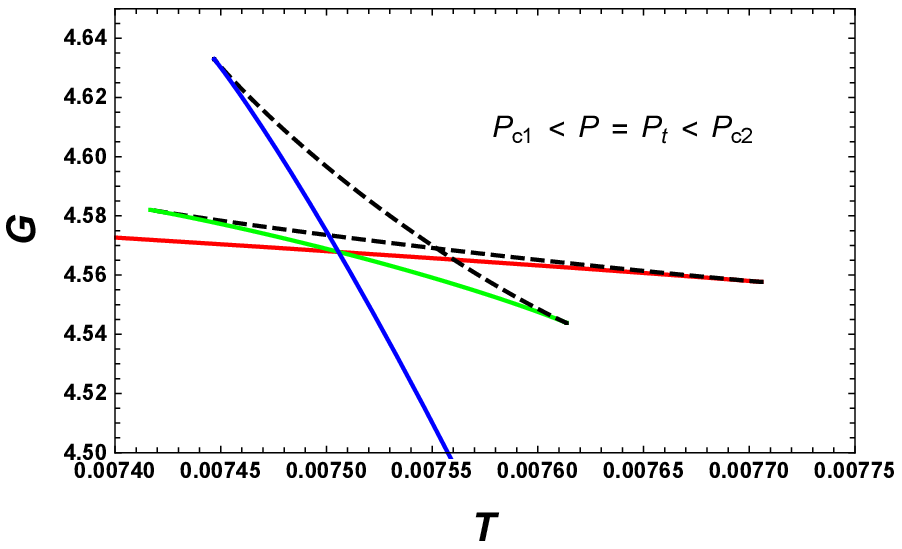}}
		\subfigure[]{
			\includegraphics[width=7cm]{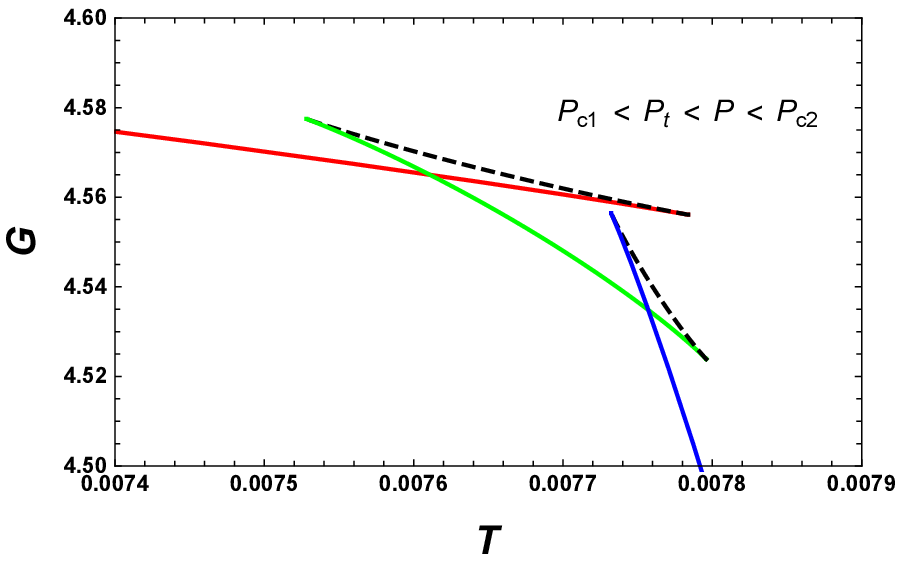}}
		\subfigure[]{
			\includegraphics[width=7cm]{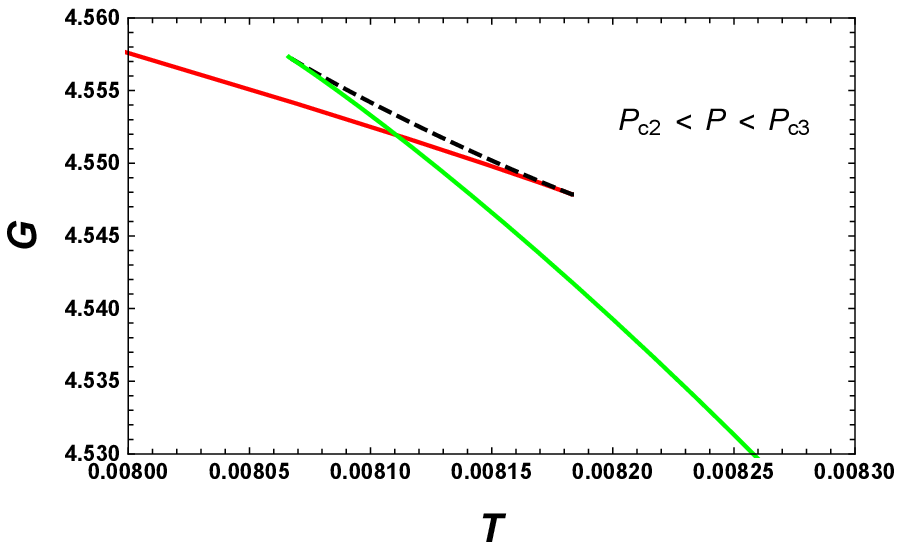}}
		\subfigure[]{
			\includegraphics[width=7cm]{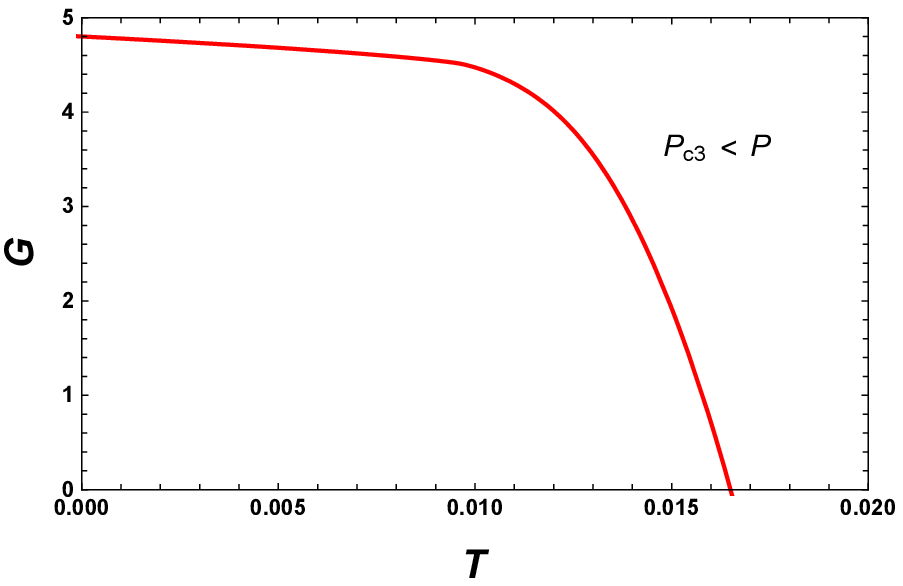}}}
	\caption{Behaviors of the Gibbs free energy $G$ with respect to the temperature $T$ with $\alpha_1=1$ and $\alpha_2=50$. The red, green, and blue curves represent small, intermediate, and large black holes, respectively. (a) $P=0.00003$. (b) $P=0.00009$. (c) $P=P_{\rm t}=0.00009608$. (d) $P=0.000105$. (e) $P=0.00015$. (f) $P=0.0003$.}\label{Fig9}
\end{figure}

For the second example, we take $\alpha_1=1$. Interestingly, we obtain three critical points by solving the condition (\ref{cond1})
\begin{eqnarray}
 T_{\rm c1}&=&0.00681279, \quad P_{\rm c1}=0.00005078, \\
 T_{\rm c2}&=&0.00802675, \quad P_{\rm c2}=0.00011537,\\
 T_{\rm c3}&=&0.00884366, \quad P_{\rm c3}=0.00022122.
\end{eqnarray}
The number of critical points indicates that there must be some phase transitions beyond the vdW-like type. To determin them, we plot the behavior of temperature $T$ with respect to the horizon radius $r_{\rm h}$ in Fig.~\ref{Fig8a}. Obviously, at some values of the pressure, there will be four extremal points for the temperature. In particular, when $P=P_{\rm t}=0.00009608$, we
construct two pairs of the equal areas described in Fig.~\ref{Fig8b}. Although these two pair areas have different values, they share the same pressure and temperature. This strongly indicates the existence of the triple point. In order to make this point more clear, we show the Gibbs free energy in Fig.\ref{Fig9}. For different cases, one or two swallowtail behaviors can be discovered. Moreover, these two characteristic swallowtail behaviors emerge, move, and disappear with the pressure. This pattern demonstrates a triple point phase structure at $P=P_{\rm t}=0.00009608$.

\begin{figure}[htb]
	\centering
	\includegraphics[width=8cm]{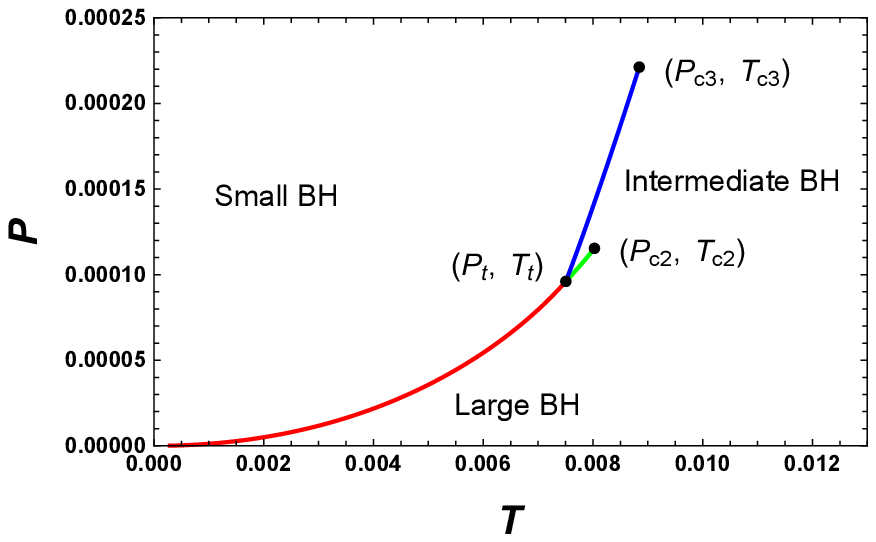}
	\caption{Phase diagram for a dyonic AdS black hole with $\alpha_1=1$ and $\alpha_2=50$.}\label{Fig10}
\end{figure}

The corresponding phase diagram is given in Fig. \ref{Fig10}, where the triple point is clearly exhibited. Another difference from the previous case is that these coexistence curves are not separate, but instead joined at the triple point.

\subsubsection{$\alpha_1=1.6$}

\begin{figure}
	\center{\subfigure[]{\label{Fig11a}
			\includegraphics[width=5cm]{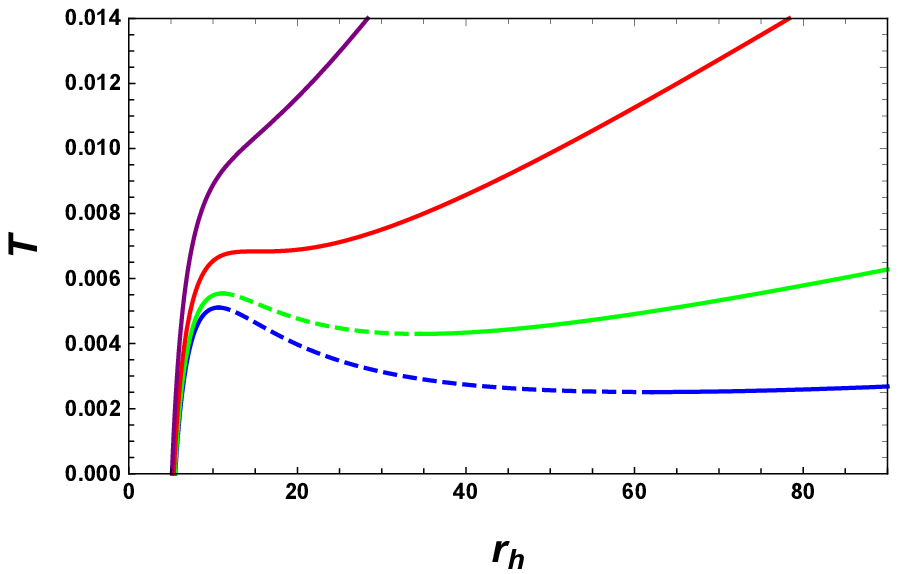}}
		\subfigure[]{\label{Fig11b}
			\includegraphics[width=5cm]{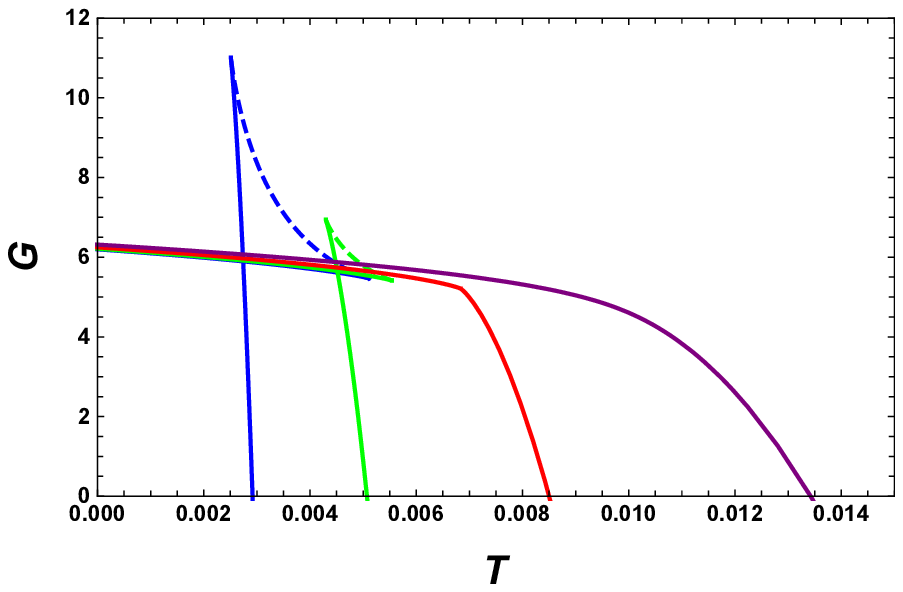}}
		\subfigure[]{\label{Fig11c}
			\includegraphics[width=5cm]{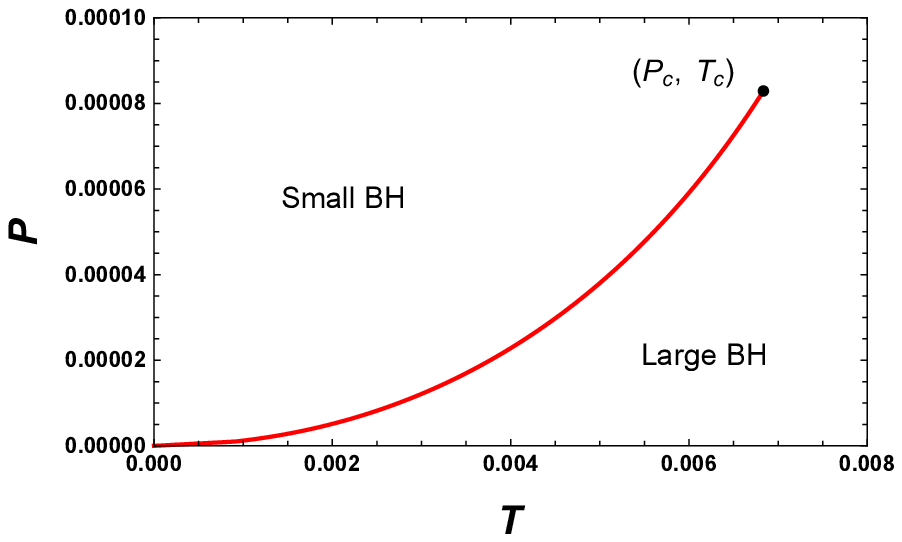}}}
\caption{(a) $T$ vs $r_{\rm h}$. (b) $G$ vs $T$. Pressures $P=0.00001$ (blue curve), $0.00003$ (green curve), $0.00008292$ (red curve), and $0.0002$ (purple curve), displayed from bottom to top in (a) and from left to right in (b). The unstable and stable branches are indicated by dashed and solid curves. (c) $P$-$T$ phase diagram. The parameters $\alpha_1=1.6$ and $\alpha_2=50$.}\label{Fig11}
\end{figure}

When $\alpha_1=1.6$, the system admits only one critical point
\begin{eqnarray}
 T_{\rm c}=0.00683463, \quad P_{\rm c}=0.00008292.
\end{eqnarray}
In this case, the behaviors of the temperature and the Gibbs free energy are shown in Figs. \ref{Fig11a} and \ref{Fig11b}. All these behaviors indicate that there is only one typical small-large black hole phase transition, which is reminiscent of the liquid/gas phase transition of the vdW fluid. Finally, the phase diagram is exhibited in Fig. \ref{Fig11c}.

\section{Critical exponents}
\label{s4}

Critical exponents describe the behavior of thermodynamical quantities near the critical points and provide a universal property of the system phase transition. In this section, we would like to numerically calculate the critical exponents for the black hole systems with quasitopological electromagnetism.

It is convenient to define the reduced thermodynamic temperature, volume, and pressure as
\begin{eqnarray}
	\tau=\frac{T}{T_{\rm c}}, \quad \nu=\frac{V}{V_{\rm c}}, \quad p=\frac{P}{P_{\rm c}}. \label{reduced}
\end{eqnarray}
Furthermore, we denote
\begin{eqnarray}
	t=\frac{T}{T_{\rm c}}-1=\tau-1, \quad \omega=\frac{V}{V_{\rm c}}-1=\nu-1. \label{para}
\end{eqnarray}
Then the critical point is shifted to $t=0$ and $\omega=0$. To calculate the critical exponents, we first obtain the corresponding equation of state by substituting the reduced parameters introduced in Eqs.~(\ref{reduced}) and (\ref{para}) into Eq.~(\ref{ppres}). Near the critical points, the reduced pressure can be expanded as
\begin{eqnarray}
	p=a_0+a_1 \omega+a_2 \omega^2+a_3 \omega^3+b_0 t +b_1 t \omega+\mathcal{O}(t\omega^2, \omega^4).
\end{eqnarray}
In our black hole systems, we find that the coefficients $a_1$ and $a_2$ vanish, and that $a_0$=1. Thus, the reduced pressure becomes
\begin{eqnarray}
	p=1+a_3 \omega^3+b_0 t +b_1 t \omega+\mathcal{O}(t\omega^2, \omega^4).\label{redpress}
\end{eqnarray}
Near the critical points we considered in the above section, we numerically obtain the values of the coefficients $a_{3}$, $b_0$, and $b_1$ by expanding the equation of state. The results are listed in Tables~\ref{tab1} and \ref{tab2}. Although these coefficients closely depend on the couplings $\alpha_1$ and $\alpha_2$, we find the universal result that $b_0$ is positive while $a_3$ and $b_1$ are negative.

\begin{table}[h]
\begin{center}
\begin{tabular}{cccccc}
\hline\hline
$\alpha_1$ & $\alpha_2$ & $a_3$ & $b_0$ & $b_1$  \\ \hline
 1 & 15 &  $-0.0461524$ & $2.6921452$ & $-0.8973817$   \\
 1 & 45 &  $-0.0369044$ & $2.7714244$ & $-0.9238081$  \\
 1 & 45 &  $-0.4079453$  & $5.1006002$ & $-1.7002001$  \\
 1 & 75 &  $-0.1533660$  & $2.8650946$ & $-0.9550316$ \\\hline\hline
\end{tabular}
\caption{Corresponding values of the expanded coefficients in Eq.~(\ref{redpress}) with a fixed $\alpha_1$.}\label{tab1}
\end{center}
\end{table}

\begin{table}[h]
	\begin{center}
		\begin{tabular}{cccccc}
			\hline\hline
			$\alpha_1$ & $\alpha_2$ & $a_3$ & $b_0$ & $b_1$  \\ \hline
			0.4 & 50 & $-0.0481278$ & $2.6764140$ & $-0.8921380$\\
			0.4 & 50 & $-0.0363386$ & $1.7104065$ & $-0.5701355$\\
			1 & 50  & $-0.0346722$  & $2.7920014$ & $-0.9306672$\\
			1 & 50  & $-0.2914387$ & $4.0208934$ & $-1.3402978$\\
			1.6 & 50 & $-0.0449899$ & $2.7030386$ & $-0.9010129$\\\hline\hline
		\end{tabular}
		\caption{Corresponding values of the expanded coefficients in Eq.~ (\ref{redpress}) with a fixed $\alpha_2$.}\label{tab2}
	\end{center}
\end{table}

Next, we attempt to calculate the critical exponents $\alpha$, $\beta$, $\gamma$, and $\delta$ by using the expanded form (\ref{redpress}).

\noindent (a) Exponent $\alpha$ describes the behavior of the specific heat at constant volume:
\begin{eqnarray}
	C_V=T\frac{\partial S}{\partial T}\bigg|_V \propto |t|^{-\alpha}.
\end{eqnarray}	
Since $S\sim V^{2/3}$, one easily obtains $C_V=0$. Thus the exponent $\alpha=0$.

\noindent (b) Exponent $\beta$ governs the behavior of the order parameter $\eta=V_{\rm l}-V_{\rm s}$ measuring the difference in thermodynamic volume between the coexistence large and small black holes for given isotherm:
\begin{eqnarray}
	\eta=V_{\rm l}-V_{\rm s} \propto |t|^\beta.
\end{eqnarray}
Considering the fact that the coexistence small and large black holes satisfy the equation of state, we have
\begin{eqnarray}
 p=1+a_3 \omega_{\rm s}^3+b_0 t +b_1 t \omega_{\rm s}=1+a_3 \omega_{\rm l}^3+b_0 t +b_1 t \omega_{\rm l}.\label{adad}
\end{eqnarray}
Differentiating the reduced pressure in Eq.~(\ref{redpress}) for a fixed $t<0$, we have
\begin{eqnarray}
	dp=(3 a_3 \omega^2 + b_1 t)d\omega.\label{dp}
\end{eqnarray}
In the reduced parameter space, it is easy to verify that the Maxwell's equal area law $\oint \omega dp=0$ holds, and it reduces to
\begin{eqnarray}
 \int_{\omega_{\rm s}}^{\omega_{\rm l}}\omega(3 a_3 \omega^2 + b_1 t)d\omega=0,\label{dopd}
\end{eqnarray}
where $\omega_{\rm s}$ and $\omega_{\rm l}$ are the reduced volumes of the small and large black holes, respectively. Solving Eqs.~(\ref{adad}) and (\ref{dopd}), we have $\omega_{\rm s}=-\omega_{\rm l}=-\sqrt{\frac{b_1}{a_3}}\sqrt{-t}$. Hence the order parameter $\eta$ is
\begin{eqnarray}
	\eta=V_{\rm c}(\omega_{\rm l}-\omega_{\rm s})=2\sqrt{\frac{b_1}{a_3}} V_{\rm c}\sqrt{-t},
\end{eqnarray}
indicating that the exponent $\beta=\frac{1}{2}$.

\noindent (c) Exponent $\gamma$ describes the behavior of the isothermal compressibility $\kappa_T$ defined as
\begin{eqnarray}
 \kappa_T=-\frac{1}{V}\frac{\partial V}{\partial P}|_T\propto |t|^{-\gamma}.
\end{eqnarray}
From the expressions of the reduced pressure in Eq.~(\ref{redpress}), one can easily get
\begin{eqnarray}
 \frac{\partial V}{\partial P}|_T=\frac{1}{b_1}\frac{V_{\rm c}}{P_{\rm c}}\frac{1}{t}+\mathcal{O}(\omega).
\end{eqnarray}
As a result,
\begin{eqnarray}
 \kappa_T=-\frac{1}{V}\frac{\partial V}{\partial P}|_T\propto-\frac{1}{b_1}\frac{V_{\rm c}}{P_{\rm c}}\frac{1}{t},
\end{eqnarray}
so one gets $\gamma=1$.

\noindent (d) Exponent $\delta$ characterizes the following behavior on the critical isotherm $T=T_{\rm c}$:
\begin{eqnarray}
	|P-P_{\rm c}|\propto|V-V_{\rm c}|^\delta.
\end{eqnarray}
Setting $t=0$ in Eq.~(\ref{redpress}), one arrives at
\begin{eqnarray}
	p-1=a_3\omega^3,
\end{eqnarray}
which leads to $\delta=3$.

In summary, the four critical exponents are obtained as
\begin{eqnarray}
	\alpha=0, \quad \beta=\frac{1}{2},  \quad \gamma=1, \quad \delta=3,
\end{eqnarray}
which is the same as that of the mean field theory and the four-dimensional charged RN-AdS black hole system~\cite{Kubiznak2012}.

\section{Discussions and conclusions}
\label{s5}

In this paper, by treating the cosmological constant as the thermodynamic pressure in the extended phase space, we studied the thermodynamics and phase transition for the dyonic AdS black hole with quasitopological electromagnetism. The results show that, compared to charged RN-AdS black holes, there are more rich black hole phase structures, such as a triple point and a separate coexistence curve.

Considering the coupling $\alpha_2$ as a new thermodynamic variable, we obtained the first law and the Smarr formula and found they are consistent with each other. By freely varying the parameters $\alpha_1$ and $\alpha_2$, respectively, we studied the phase transition and phase diagram via the behaviors of the temperature and Gibbs free energy along the isobaric curves.

We first considered the case in which $\alpha_1$=1, while $\alpha_2$ was 15, 45, or 75. When $\alpha_2=15$, we observed the small-large black hole phase transition. The phase diagrams are similar to that of the vdW fluid. For $\alpha_2=45$, four extremal points of the temperature and two swallowtail behaviors of the Gibbs free energy along certain isobaric curves can be found. These imply rich phase structures. Detailed study shows that there is a triple point at which three black hole phases coexist, just like that of water with coexisting ice, liquid, and vapor phases. This phase diagram has not been found in charged RN-AdS black holes. And thus it is a novel phenomenon for black holes with quasitopological electromagnetism. When the coupling is further increased such that $\alpha_2=75$, we observed only one swallowtail behavior, so only the small-large black hole phase transition is present in this case.

Then we varied $\alpha_1$ while instead keeping $\alpha_2$ fixed at $50$. For $\alpha_1=0.4$, we found a novel phase structure. Although in these cases, there are two swallowtail behaviors, some parts of them fall in the negative temperature regions, indicating that they are unphysical. After detailed study, we discovered that there are two separate coexistence curves in the phase diagrams. One starts at the origin and ends at a critical point, while the other one starts at a certain point with vanishing temperature, then extends to another critical point. This is also a novel diagram for a dyonic AdS black hole with quasitopological electromagnetism. When $\alpha_1$ is increased such that $\alpha_1=1$, a characteristic triple point is observed. When $\alpha_1$ approaches 1.6, there is only a small-large black hole phase transition of the vdW-like type.

After obtaining the phase diagrams, we calculated the critical exponents for each case. Employed with the expansion of the pressure near the critical points, we obtained the exponents $\alpha=0$, $\beta=\frac{1}{2}$, $\gamma=1$, and $\delta=3$, which display the same values as in mean field theory.

These results uncover intriguing thermodynamic properties and phase diagrams for dyonic AdS black holes with quasitopological electromagnetism. The study is also worth generalizing to higher-dimensional cases. Furthermore, black hole microstructure still remains to be tested by constructing the Ruppeiner geometry. These are valuable to further disclose the influence of quasitopological electromagnetism on the black hole thermodynamics. It is also worth further examining the emergent new black hole phases on the dynamical stability of the black holes, as well as in the dual field theories according to the AdS/CFT. Furthermore, since there are two unstable photon spheres, novel features imprinted in the black hole shadow should be uncovered~\cite{Gralla}.

\section*{ACKNOWLEDGMENTS}
This work was supported by the National Natural Science Foundation of China (Grants No. 12075103, No. 11675064, and No. 12047501) and the 111 Project (Grant No. B20063).

\end{document}